\documentclass[final,3p,times]{elsarticle}

\usepackage{amssymb}

\usepackage[utf8]{inputenc}
\usepackage{subcaption}
\usepackage{caption}
\usepackage{xcolor}
\usepackage{amsmath}
\usepackage{float}

\newcommand{\intL}{\int_{\cal L}}

\begin{document}

\newcommand{\beq}{\begin{equation}}
\newcommand{\beql}[1]{\begin{equation}\label{#1}}
\newcommand{\eeq}{\end{equation}}
\newcommand{\bea}{\begin{eqnarray}}
\newcommand{\beal}[1]{\begin{eqnarray}\label{#1}}
\newcommand{\eea}{\end{eqnarray}}
\newcommand{\bean}{\begin{eqnarray*}}
\newcommand{\eean}{\end{eqnarray*}}

\newcommand{\sig}{\sigma}
\newcommand{\eps}{\varepsilon}
\newcommand{\del}{\delta}

\newcommand{\mbf}[1]{\mathbf{#1}}
\newcommand{\sbf}[1]{\boldsymbol{#1}}
\newcommand{\ignore}[1]{}

\newcommand{\beps}{\sbf{\varepsilon}}
\newcommand{\bsig}{\sbf{\sigma}}
\newcommand{\blam}{\sbf{\lambda}}
\newcommand{\bmu}{\sbf{\mu}}
\newcommand{\ba}{\mbf{a}}
\newcommand{\bb}{\mbf{b}}
\newcommand{\bi}{\mbf{i}}
\newcommand{\bB}{\mbf{B}}
\newcommand{\bS}{\mbf{S}}
\newcommand{\bV}{\mbf{V}}
\newcommand{\bA}{\mbf{A}}
\newcommand{\bE}{\mbf{E}}
\newcommand{\bU}{\mbf{U}}
\newcommand{\bR}{\mbf{R}}
\newcommand{\bX}{\mbf{X}}
\newcommand{\bK}{\mbf{K}}
\newcommand{\bd}{\mbf{d}}
\newcommand{\bs}{\mbf{s}}
\newcommand{\be}{\mbf{e}}
\newcommand{\bff}{\mbf{f}}
\newcommand{\bF}{\mbf{F}}
\newcommand{\bg}{\mbf{g}}
\newcommand{\bG}{\mbf{G}}
\newcommand{\bL}{\mbf{L}}
\newcommand{\bM}{\mbf{M}}
\newcommand{\bH}{\mbf{H}}
\newcommand{\bkron}{\mbf{1}}
\newcommand{\sigm}{\sigma_{\rm m}}
\newcommand{\epsm}{\varepsilon_{\rm m}}
\newcommand{\bI}{\mbf{I}}
\newcommand{\bIv}{\mbf{I}_{\rm K}}
\newcommand{\bId}{\mbf{I}_{\rm D}}
\newcommand{\bIs}{\mbf{I}_{\rm S}}
\newcommand{\bCe}{\mbf{C}_{\rm e}}
\newcommand{\bDe}{\mbf{D}_{\rm e}}
\newcommand{\bC}{\mbf{C}}
\newcommand{\bD}{\mbf{D}}
\newcommand{\bu}{\mbf{u}}
\newcommand{\bv}{\mbf{v}}
\newcommand{\bw}{\mbf{w}}
\newcommand{\bz}{\mbf{z}}
\newcommand{\bT}{\mbf{T}}
\newcommand{\bt}{\mbf{t}}
\newcommand{\bn}{\mbf{n}}
\newcommand{\bN}{\mbf{N}}
\newcommand{\bx}{\mbf{x}}
\newcommand{\bq}{\mbf{q}}
\newcommand{\bh}{\mbf{h}}

\newcommand{\bnab}{\sbf{\nabla}}
\newcommand{\bnabs}{\sbf{\nabla}_{\rm s}}

\newcommand{\third}{\mbox{$\frac 1 3$}}
\newcommand{\half}{\mbox{$\frac 1 2$}}
\newcommand{\dV}{\,\mbox{d}V}
\newcommand{\dx}{\,\mbox{d}x}
\newcommand{\dxi}{\,\mbox{d}\xi}
\newcommand{\deta}{\,{\rm d}\eta}
\newcommand{\dS}{\,\mbox{d}S}
\newcommand{\ds}{\,\mbox{d}s}
\newcommand{\dA}{\,\mbox{d}A}
\newcommand{\dt}{\,\mbox{d}t}
\newcommand{\pard}[2]{\frac{\partial #1}{\partial #2}}
\newcommand{\parder}[2]{\frac{\partial #1}{\partial #2}}

\begin{frontmatter}



\title{Integral Micromorphic Model for Band Gap in~1D~Continuum}


\author[1]{Milan Jir\'{a}sek}
\ead{milan.jirasek@fsv.cvut.cz}

\author[2,3]{Martin Hor\'{a}k}
\ead{martin.horak@cvut.cz}

\author[4]{Michal \v{S}mejkal\corref{cor1}}
\ead{michal.smejkal@fsv.cvut.cz}

\cortext[cor1]{Corresponding author}
\address[1,2,4]{Czech Technical University in Prague, Faculty of Civil Engineering, Department of Mechanics, Thákurova 2077/7, 166 29 Prague 6, Czech Republic}
\address[3]{Czech Academy of Sciences, Institute of Information Theory and Automation, Pod vod\'{a}renskou
v\v{e}\v{z}\'{\i}~4, 182~00~Praha~8, Czech Republic}

\begin{abstract}
The design of band-gap metamaterials, i.e., metamaterials with the capability to inhibit wave propagation of a specific frequency range, has numerous potential engineering applications, such as acoustic filters and vibration isolation control. In order to describe the behavior of such materials, a novel integral micromorphic elastic continuum is introduced, and its ability to describe band gaps is studied in the one-dimensional setting. The nonlocal formulation is based on a modification of two terms in the expression for potential energy density. The corresponding dispersion equation is derived and converted to a dimensionless format, so that the effect of individual parameters can be 
described in the most efficient way. The results indicate
that both suggested nonlocal modifications play an important role. 
The original local micromorphic model reproduces a band gap 
only in the special, somewhat artificial case, when
the stiffness coefficient associated with the gradient of the micromorphic variable vanishes. On the other hand, 
the nonlocal formulation can provide band gaps even for
nonzero values of this coefficient, provided that the
penalty coefficient that enforces coupling between the micromorphic variable and nonlocal strain is sufficiently high and the micromorphic stiffness is sufficiently low.
\end{abstract}




\end{frontmatter}


\section{Introduction}

In recent years, mechanical metamaterials have attracted attention due to the possibility of designing materials with unusual and tailored properties. A typical mechanical metamaterial is composed of a pattern of repeating unit cells. Thus, the interest in metamaterials has increased hand in hand with the development of 3D printing techniques that enable the production of metamaterials in an effective and reproducible way \cite{nevzerka2018jigsaw,lei20193d}. One widely studied metamaterial type is the so-called Locally Resonant Acoustic Metamaterial (LRAM) \cite{craster2012acoustic, ma2016acoustic,sugino2016mechanism}. The dispersion curve of such material shows multiple branches (called acoustic and optical) separated by a band gap. Therefore, LRAM can attenuate acoustic waves within a specific frequency range, leading to numerous potential engineering applications such as acoustic cloaking, acoustic filters, and vibration isolation control \cite{norris2015acoustic, miniaci2016large}. 

The design of the metamaterials relies upon a robust and efficient computational tool allowing virtual simulations of material behavior and properties. The straightforward technique is a direct numerical simulation (usually based on the finite element method), considering the complete microstructure. However, direct numerical simulations of large metamaterial structures can quickly become unfeasible due to significant computational costs stemming from the representation of all geometric details of each unit cell, rendering them impractical for engineering practice. Therefore, homogenization techniques to capture the behavior of Locally Resonant Acoustic Metamaterials have been developed, see, e.g., \cite{roca2019computational,liu2021computational,van2019transient}.

Besides homogenization methods, various phenomenological theories, including extended continuum theories such as higher-gradient \cite{askes2008four, fish2002non} and nonlocal elasticity \cite{eringen1972linear, lim2015higher}, have been proposed as approaches that can incorporate the influence of material microstructure on the macroscopic behavior, e.g., on wave propagation. A comprehensive investigation of dispersion properties of various integral-type and gradient-type elasticity theories was presented in \cite{jirasek2004nonlocal}. Even though the gradient and integral continuum enrichments allow to model dispersive behavior,  their dispersive diagrams contain only one branch. A possible approach to capturing acoustic as well as optical branches of the dispersion diagram is to resort to micromorphic continua \cite{eringen1964nonlinear,mindlin}. In comparison to gradient-enhanced models, micromorphic continuum introduces additional degrees of freedom describing the deformation of the microstructure. Nonetheless, the classical micromorphic continua, including the well-known Cosserat continuum \cite{cosserat1909theorie,herrmann1968applications}, microstretch continuum \cite{eringen1990theory, tomar2006propagation}, and the full micromorphic continuum \cite{eringen1964nonlinear, berezovski2013dispersive}, cannot capture the band gap, as was shown in \cite{madeo2016complete}. Moreover, a detailed analysis of the wave propagation in micromorphic continua in one dimension was
presented in \cite{berezovski2013dispersive} and did not report any band gap.

A novel relaxed micromorphic continuum capable of capturing the band gap has been proposed recently in \cite{neff2014unifying}, and its properties have been studied in a series of papers, see, e.g., \cite{ghiba2015relaxed, madeo2015band, neff2017real, madeo2017role, d2020effective, ghiba2021existence}. Note that in the one-dimensional case, the micromorphic effects of the relaxed micromorphic continuum vanish since they describe relative rotations of the microstructure with respect to the macroscopic matrix. In \cite{madeo2016complete}, the authors of the relaxed micromorphic continuum claimed that the relaxed micromorphic model was the only nonlocal continuum model able to account for the description of band gaps. Nonetheless,  the same authors later showed in \cite{madeo2017role} that by adding a term containing mixed temporal and spatial derivatives of displacement, classical micromorphic models could account for the band gap. A micromorphic model with mixed temporal-spatial derivatives of the micromorphic variable, capable of modeling band gaps, was developed from microstructural arguments in \cite{nejadsadeghi2020role}. Still, the relaxed micromorphic continuum remains the only nonlocal enriched model able to describe band gaps without considering enrichments of the kinetic energy expression by terms with mixed temporal and spatial derivatives of some variable.

In this paper, we develop a new nonlocal model for modeling of band gaps. The formulation of the model is motivated by results presented in \cite{jirasek2004nonlocal}, showing that the integral formulation covers several proposed gradient models such as \cite{fish2002non, metrikine2002one} as special cases and is also capable of reproducing the dispersion curve of a discrete lattice merely by choosing an appropriate nonlocal kernel. To simulate band gaps, we propose to further enhance the integral model by micromorphic degrees of freedom, leading to a new integral micromorphic formulation.  
The point of departure for the micromorphic formulation is the specification of the free energy density $\psi$ (see, e.g., \cite{forest2016nonlinear}), which is in the 1D case written as
\begin{equation}\label{eq_psi}
\psi = \frac{1}{2}E\varepsilon^2 + \frac{1}{2}A \chi'^2 + \frac{1}{2}H(\varphi(\varepsilon)-\chi)^2
\end{equation}
where $E$ is the elastic modulus $A$, and $H$ are additional parameters, and the prime denotes differentiation with respect to the spatial coordinate, $x$. The displacement, $u$, is linked to the local strain, $\eps$, by the standard relation $\eps=u'$.
In addition to the classical quadratic function of the strain $\varepsilon$, the generalized expression (\ref{eq_psi}) contains quadratic terms with the gradient of the micromorphic variable $\chi$, and with a generalized relative strain variable $e = \varphi(\varepsilon)-\chi $ providing coupling between the strain and the micromorphic variable. Note that the meaning of the micromorphic variable is defined by the selection of function $\varphi(\varepsilon)$ and can in general lead to various types of micromorphic continua, such as Cosserat continua and microstretch continua. However, in one dimension, the micromorphic effect either disappears (Cosserat or relaxed micromorphic continua), or the formulation of the micromorphic continua is reduced to the form presented in Equation (\ref{eq_psi}). 
In this paper, a novel integral micromorphic continuum is developed, for which $\varphi(\varepsilon)$ is chosen as the averaged (nonlocal) strain. In the context of the nonlocal integral continuum, the first term in (\ref{eq_psi}) is also enhanced such that it includes spatial averaging. As will be shown, the resulting formulation allows modeling of the band gap in the dispersion curve.

The paper is structured as follows. Firstly, the equations of motion for the proposed integral micromorphic continuum are derived using the Hamilton variational principle. Subsequently, assuming  harmonic wave propagation, the dispersion equation is obtained and transformed to the dimensionless form. Afterward, the effects of the individual model parameters as well as the influence of the integral averaging on the shape of the dispersion curve are examined. For simplicity and in order to allow for analytical derivations as much as possible, only a 1D problem is studied in the paper.   

\section{One-dimensional Integral Micromorphic Model}\label{sec:2}

\subsection{Equations of motion}\label{sec5.1}

The free energy density function for the local micromorphic model introduced in equation \eqref{eq_psi} is now generalized to include nonlocal effects both in the coupling term $\varphi(\varepsilon)$ and in the strain-related term. The generalized expression reads 
\begin{equation} \label{eq_free_en}
\psi(x) = \frac{1}{2}\varepsilon(x) \intL E\alpha_0(x,\xi)\varepsilon(\xi) \;{\rm d}\xi + \frac{1}{2}A \chi'^2(x) + \frac{1}{2}H\left(\intL \alpha_1(x,\xi)\varepsilon(\xi) \;{\rm d}\xi-\chi(x)\right)^2
\end{equation}
where ${\cal L}$ is the one-dimensional domain representing the analyzed body (here we consider ${\cal L}=(-\infty,\infty)$), 
$\alpha_0$ and $\alpha_1$ are nonlocal averaging kernels and $A$ is the generalized micromorphic modulus. 
Parameter $H$ can be interpreted as a penalty coefficient; in the limit case when $H \to \infty$, the micromorphic variable corresponds to the nonlocal strain
evaluated with weight function $\alpha_1$, and the presented model simplifies to the higher-order nonlocal strain gradient elasticity which was introduced in \cite{jirasek2004nonlocal} and studied in \cite{lim2015higher}.  
For simplicity, we do not mark explicitly
that the state variables in general depend
not only on the spatial coordinate but also on time, $t$.

It would also be possible to introduce spatial averaging for inertia, however, this modification is not considered here, and the kinetic energy is generalized only in the micromorphic sense, i.e., its density is given by
\begin{equation}\label{eq3}
{\cal E}_{kin} = \frac{1}{2}\rho\dot{u}^2 + \frac{1}{2}\eta\dot{\chi}^2
\end{equation}
where  $\rho$ is the standard mass density and $\eta$ is a nonstandard, micromorphic density.
The superimposed dot denotes a derivative
with respect to time.

In the spirit of the Hamilton principle, we define the action functional
\beq 
{\cal S}=\int_{t_1}^{t_2} \intL \left[{\cal E}_{kin}(x,t)-\psi(x,t) \right]\dx\dt
\eeq 
and set its first variation to zero, assuming that the state of the system at times $t_1$ and $t_2$ is fixed. Substituting from (\ref{eq_free_en})--(\ref{eq3}) with $\eps$ replaced by $u'$,
evaluating the variation, integrating by parts with respect to time and taking into account that the variations at $t_1$ and $t_2$ vanish, we obtain the stationarity condition
\beq \label{eq5}
-\intL\left[\rho\ddot{u}(x)\delta u(x)+\eta\ddot{\chi}(x)\delta\chi(x)+  \Sigma(x) \delta u'(x)\dx + \Theta_0(x)  \delta\chi(x)\dx+  \Theta_1(x)  \delta\chi'(x)\right]\dx = 0
\eeq 
in which
\bea \label{eq6}
\Sigma(x) &=& E\intL \half\left(\alpha_0(x,\xi)+\alpha_0(\xi,x)\right)u'(\xi)  \dxi + H\intL \left(\intL\alpha_1(\eta,\xi)u'(\xi)\dxi-\chi(\eta)\right)\alpha_1(\eta,x)\deta
\\
\Theta_1(x) &=& A \chi'(x)
\\
\Theta_0(x) &=& - H\left(\intL\alpha_1(x,\xi)u'(\xi)\dxi-\chi(x)\right)
\eea
can be interpreted as the stress, the higher-order stress, and the relative stress.

Condition (\ref{eq5})
should be satisfied for all admissible
variations $\delta u$ and $\delta\chi$.
Integrating the terms with $\delta u'$ and $\delta \chi'$ by parts with respect to space, we end up with
Euler-Lagrange equations
\bea \label{eq9}
\rho \ddot{u} &=& \Sigma'
\\ \label{eq10}
\eta\ddot{\chi} &=& \Theta_1' - 
\Theta_0
\eea 

Note that expression (\ref{eq6}) for $\Sigma$ contains a double integral,
but integration with respect to $\eta$ can be done in advance because it does not involve the variable fields.
Denoting
\beq \label{eq11}
E_H(x,\xi)= E\,\frac{\alpha_0(x,\xi)+\alpha_0(\xi,x)}{2}+H \intL\alpha_1(\eta,x)\alpha_1(\eta,\xi)\deta
\eeq 
we can rewrite (\ref{eq6}) as
\beq 
\Sigma(x) = \intL E_H(x,\xi)u'(\xi)  \dxi 
- H\intL \alpha_1(\xi,x)\chi(\xi)\dxi
\eeq 
The resulting equations of motion deduced from (\ref{eq9})--(\ref{eq10}) read
\bea 
\rho \ddot{u}(x,t) &=& \left(\intL E_H(x,\xi)u'(\xi,t)  \dxi 
- H\intL \alpha_1(\xi,x)\chi(\xi,t)\dxi\right)' \label{eq.1}
\\
\eta\ddot{\chi}(x,t) &=& A \chi''(x,t) +H\left(\intL\alpha_1(x,\xi)u'(\xi,t)\dxi-\chi(x,t)\right) \label{eq.4}
\eea 

Function 
$E_H$ defined in (\ref{eq11})
is a modified weight function, which turns out to be symmetric
with respect to $x$ and $\xi$ even if the original functions
$E(x,\xi)$ and $\alpha_1(x,\xi)$ are not.
If this weight function depends only on the distance, i.e., if $E_H(x,\xi)=E_0(x-\xi)$,
then the ``outer derivative'' on the right-hand side of (\ref{eq.1}) can be ``shifted'' 
into the nonlocal integrals (for simplicity, we drop the dependence on the time variable):
\bea \nonumber
\left[\intL E_0(x-\xi)u'(\xi)  \dxi \right]'&=&\intL E_0'(x-\xi)u'(\xi)  \dxi
= -\intL E_0'(\xi-x)u'(\xi)  \dxi =
\\  &=&
-\left[E_0(x-\xi)u'(\xi)\right]_{\xi=-\infty}^\infty + \intL E_0(x-\xi)u''(\xi)  \dxi = \intL E_0(x-\xi)u''(\xi)  \dxi 
\eea 
Here we have used certain assumed properties of the weight function, e.g., $E_0(r)\to 0$ as $|r|\to\infty$.   
If the weight function $\alpha_1$ also depends just on the distance, one can show by an analogous procedure that
\beq 
\left[\intL \alpha_1(\xi,x)\chi(\xi)\dxi\right]' = \intL \alpha_1(\xi,x)\chi'(\xi)\dxi
\eeq 
and equation \eqref{eq.1} can then be rewritten as
\beq \label{eq.3}
\rho \ddot{u}(x,t) = \intL E_H(x,\xi)u''(\xi,t)  \dxi 
- H\intL \alpha_1(x,\xi)\chi'(\xi,t)\dxi
\eeq

\subsection{Dispersion relation}
The dispersion equation, i.e., the relation between circular frequency $\omega$ and wave number $k$, is now derived by assuming harmonic propagation of both displacement and micromorphic variable. For an infinite
and macroscopically homogeneous body, it is natural to
expect that the weight functions depend on distance only.
With a slight abuse of notation, we will write
$\alpha_j(x-\xi)$ instead of $\alpha_j(x,\xi)$, $j=0,1$,
and also $E_H(x-\xi)$ instead of $E_H(x,\xi)$.
Plugging the harmonic ansatz
 \begin{eqnarray}
u(x,t) &=& U{\rm e}^{i(kx-\omega t)}
\\
\chi(x,t) &=& X{\rm e}^{i(kx-\omega t)}
\end{eqnarray}
into equations \eqref{eq.3} and \eqref{eq.4} and multiplying both equations by ${\rm e}^{-i(kx-\omega t)}$ leads to conditions
\begin{eqnarray}
\nonumber
-\rho \omega^2U &=& -k^2U\intL E_H(r){\rm e}^{-ikr} {\rm d} r  -  i k H X
\intL \alpha_1(r){\rm e}^{-ikr} \;{\rm d} r
\\
\nonumber
-\eta \omega^2X &=& -Ak^2X + H\left( ik U\intL \alpha_1(r){\rm e}^{-ikr} \;{\rm d}r -X \right)
\end{eqnarray}
which can be written in a matrix form as
\beq\label{eq.2}
\begin{bmatrix}
\rho\omega^2 -k^2(E\alpha_0^*(k)+H\alpha_1^{*2}(k)) & -ik H\alpha_1^*(k)
\\
ikH\alpha_1^*(k) &  \eta \omega^2-k^2A-H
\end{bmatrix} \begin{Bmatrix}
U
\\
X
\end{Bmatrix} = \begin{Bmatrix}
0
\\
0
\end{Bmatrix}
\eeq
where $\alpha^*_j(k) = \intL \alpha_j(r){\rm e}^{-ikr}{\rm d}r$, $j=0,1$, denotes the Fourier transform of the $j$-th nonlocal kernel.
It is worth noting that since ${\cal L}=(-\infty,\infty)$ and the weight function
$\alpha_j$ is always even, the imaginary part 
of $\alpha_j^*$ vanishes. Therefore, $\alpha_j^*$ can also be defined by the
cosine Fourier transform
\beq 
\alpha^*_j(k) = \intL \alpha_j(r)\cos kr\,{\rm d}r
\eeq 

A nontrivial solution of homogeneous linear algebraic equations (\ref{eq.2}) exists
if and only if the determinant of the matrix  vanishes. This condition results in the dispersion equation
\beq 
\rho\eta\omega^4-[\rho(k^2A+H)+\eta k^2C^*(k)]\omega^2
+k^4AC^*(k)+k^2EH\alpha_0^*(k)=0
\eeq
where $C^*(k)=E\alpha_0^*(k)+H\alpha_1^{*2}(k)$. This is a quadratic equation in terms of $\omega^2$ and its solution reads
\beq\label{eq:dispnlmm} 
\omega^2 = \frac{\rho(k^2A+H)+\eta k^2C^*(k)\pm\sqrt{[\rho(k^2A+H)-\eta k^2C^*(k)]^2+4\rho\eta H^2k^2 \alpha_1^{*2}(k)}}{2\rho\eta} 
\eeq 
For $C^*(k)=E+H$ and $\alpha_1^*(k)=1$, this reduces to the dispersion relation 
\begin{equation}
\omega^2 =  \frac{k^2((E+H)\eta+A\rho)+H\rho}{2\rho \eta} \pm \frac{\sqrt{\left[k^2((E+H)\eta-A\rho)-H\rho\right]^2 + 4k^2H^2\rho\eta}}{2\rho\eta}  
\end{equation}
valid for the ``local'' micromorphic model,
i.e., for the model with
the free energy density given by (\ref{eq_psi}) with $\varphi(\eps)\equiv\eps$.

\subsection{Dimensionless formulation}

Since we are dealing with 5 model parameters ($E$, $A$, $H$, $\rho$, and $\eta$)
that have 3 independent units, we can reduce the description
to $5-3=2$ dimensionless input parameters. Our choice here
is to consider $\kappa=H/E$ and $\lambda^2=A\rho/(E\eta)$.
Therefore, in the dimensionless format, $\rho$, $\eta$,
and $E$ are replaced by 1, $H$ is replaced by $\kappa$, and $A$ by $\lambda^2$. 
The wave number $k$ and circular frequency $\omega$
will be transformed into their dimensionless counterparts $\tilde{k}=k\sqrt{\eta/\rho}$ and
$\tilde\omega=\omega \sqrt{\eta/E}$. 
The physical meaning of parameter $\lambda$ will become clear later (it is the ratio of the speed of fictitious ``purely micromorphic'' waves that would appear in a fully decoupled local model with
$H=0$ and the standard elastic wave speed).

In the one-dimensional setting, the kernels $\alpha_j$ have units 1/m, and so we can convert them into dimensionless kernels $\tilde\alpha_j=\alpha_j\sqrt{\eta/\rho}$ and
perform averaging by integrating with respect to
the dimensionless spatial coordinate $\tilde{x}=x\sqrt{\rho/\eta}$.
The Fourier image $\alpha_j^*$ is dimensionless
but it is considered as function of the wave number $k$, which has units 1/m. Therefore, we will introduce the symbol $\tilde\alpha_j^*$ for the
Fourier transform as function of the dimensionless
wave number $\tilde k$. To explain that better, 
let us look at the following transformation
(for convenience, we introduce an auxiliary length parameter $\ell=\sqrt{\eta/\rho}$):
\bea
\alpha^*(k)=\intL\alpha(x){\rm e}^{-ikx}\dx =
\intL \alpha(\ell\tilde x){\rm e}^{-ik\ell\tilde x}\ell\,{\rm d}\tilde x=
\intL \ell\alpha(\ell\tilde x){\rm e}^{-i\tilde{k}\tilde x}\,{\rm d}\tilde x
=
\intL \tilde\alpha(\tilde x){\rm e}^{-i\tilde{k}\tilde x}\,{\rm d}\tilde x = \tilde\alpha^*(\tilde k)
\eea

It should be noted that the definition of the kernel typically involves at least one length-scale parameter. For instance, the Gaussian kernel
\beq \label{eq:gauss1}
\alpha(x)=\frac{1}{a\sqrt{\pi}}{\rm e}^{-x^2/a^2}
\eeq 
has length parameter $a$, and the corresponding
Fourier transform is
\beq \label{eq:gauss2}
\alpha^*(k) = {\rm e}^{-k^2a^2/4}
\eeq 
The dimensionless counterparts
\bea
\tilde\alpha(\tilde x)&=&\frac{\ell}{a\sqrt{\pi}}{\rm e}^{-\ell^2\tilde{x}^2/a^2}
=\frac{1}{\tilde{a}\sqrt{\pi}}{\rm e}^{-\tilde{x}^2/\tilde{a}^2}
\\
\tilde\alpha^*(\tilde k) &=& {\rm e}^{-\tilde{k}^2a^2/(4\ell^2)}=
{\rm e}^{-\tilde{k}^2\tilde{a}^2/4}
\eea
contain a dimensionless parameter $\tilde{a}=a/\ell$ where
$\ell=\sqrt{\eta/\rho}$ is the length scale already set by other
parameters of the model, $\eta$ and $\rho$.
Now it is easy to rewrite the dispersion relation (\ref{eq:dispnlmm})
in the dimensionless form 
\beq\label{eq:dispnlmm3} 
2\tilde\omega^2 = \kappa+ \tilde{k}^2(\lambda^2+\tilde\alpha_0^*(\tilde k)+\kappa\tilde\alpha_1^{*2}(\tilde k))\pm\sqrt{[\kappa+ \tilde{k}^2(\lambda^2-\tilde\alpha_0^*(\tilde k)-\kappa\tilde\alpha_1^{*2}(\tilde k))]^2+4\kappa^2\tilde{k}^2 \tilde\alpha_1^{*2}(\tilde k)}
\eeq
The positive sign before the square root corresponds to the optical branch and the negative one to the acoustic branch.

\section{Results and Discussion}
The shape of the dispersion diagram is controlled by
parameters $\kappa$ and $\lambda$ and, in the nonlocal case, also by two additional dimensionless parameters used in the expressions for Fourier transforms $\tilde\alpha^*_0$ and $\tilde\alpha^*_1$ (and of course by the choice of a specific type of the kernels, e.g., Gaussian).
The effect of the individual model parameters is now investigated. The local micromorphic model is studied first. 

\subsection{Local micromorphic model}\label{sec:3.1}

The local dimensionless formulation can be recovered from the nonlocal model by setting both averaging kernels $\tilde\alpha_0(\tilde x)$ and $\tilde\alpha_1(\tilde x)$ to the Dirac delta distribution. Consequently, the associated Fourier images $\tilde\alpha^*_0(\tilde k)$ and $\tilde\alpha^*_1(\tilde k)$ are set to 1 and the local counterpart to equation \eqref{eq:dispnlmm3} reads
\beq\label{eq:dispnlmm4} 
2\tilde\omega^2 = \kappa+ \tilde{k}^2(\lambda^2+1+\kappa)\pm\sqrt{[\kappa+ \tilde{k}^2(\lambda^2-1-\kappa)]^2+4\kappa^2\tilde{k}^2 }
\eeq
For $\tilde k=0$, i.e., in the long-wave limit, the corresponding frequency is given by
\beq 
\tilde\omega_0 = \sqrt{\frac{\kappa \pm\sqrt{\kappa^2}}{2}}
\eeq 
which means that the frequency
on the acoustic branch is 0 and on the optical branch it is $\sqrt{\kappa}$. Therefore, parameter $\kappa$ controls the initial
value on the  optical branch. This is also valid for the general nonlocal model because $\tilde\alpha^*(0)=1$
for any weight function satisfying the normalizing 
condition $\intL\alpha(\xi)\dxi=1$.

In the opposite extreme case, for $\tilde k\to \infty$, i.e.,
in the short-wave limit, the dispersion relation can be approximated by
\beq 
\tilde\omega \approx \tilde k\, \sqrt{\frac{\lambda^2+1+\kappa \pm\left(\lambda^2-1-\kappa\right)}{2}}
\eeq 
This means that the optical branch asymptotically approaches a straight line of slope $\max(\lambda,\sqrt{1+\kappa})$ and the acoustic branch a straight line of slope $\min(\lambda,\sqrt{1+\kappa})$.
This terminal slope represents the dimensionless speed of very short waves
(i.e., the actual speed normalized by the standard elastic wave speed, $c_0=\sqrt{E/\rho}$). On the other hand, the dimensionless speed of very long acoustic waves
is represented by the initial slope of the acoustic branch
and is always equal to 1, which corresponds to standard
elasticity. This is natural, because if the wavelength
is much larger than the internal length of the enriched
continuum, the enrichment effect becomes negligible.

The effect of parameter $\kappa$ is illustrated
by the dispersion curves in Figure \ref{fig_local_kappa}. 
In the theoretical limit case when $\kappa=0$ (i.e., when $H=0$), equation \eqref{eq:dispnlmm4} yields
\beq\label{eq:dispnlmm5} 
\tilde\omega =  \tilde{k}\,\sqrt{\frac{\lambda^2+1\pm \left|\lambda^2-1\right|}{2}}
\eeq
This indicates that if the coupling between the micromorphic variable  and the strain is not enforced (i.e., the penalty parameter $H$ is set to zero), the two branches of the dimensionless dispersion curve reduce to two straight lines with slopes 1 and $\lambda$, resp., both starting at the origin. The branch with slope 1 corresponds to the standard elasticity theory,
and the branch with slope $\lambda$ is 
artificial, because the micromorphic variable
is for $H=0$ totally independent of displacement or strain. However, the artificial branch 
indicates what can be expected
for the optical branch in the case of weak coupling with a small but nonzero value of
penalty parameter $H$ (note the blue curves
with hollow circles, corresponding to $\kappa=0.2$).
Also, this analysis confirms the previously announced
interpretation of parameter $\lambda$ as the ratio 
between two wave speeds---one is
the speed $\sqrt{A/\eta}$ of fictitious
``micromorphic waves'' that would arise for a model
with no coupling between the micromorphic variable and
strain, and the other is the standard elastic wave speed,
$\sqrt{E/\rho}$.

In the opposite limit case when $\kappa\to\infty$ (i.e., $H\to\infty$, resulting into a strongly coupled model), the micromorphic variable $\chi$ coincides with the strain $\varepsilon$ and the model reduces to strain gradient elasticity with higher-order inertia, which was introduced in \cite{metrikine2002one} by Metrikine \& Askes. The corresponding dimensionless dispersion relation  reads
\beq\label{eq:dispnlmm6} 
\tilde\omega =\tilde k\, \sqrt{ \frac{1+\lambda^2 \tilde k^2}{1+\tilde k^2} }
\eeq
and in Figure~\ref{fig_local_kappa} it is plotted in brown. The optical branch in
this limit case disappears---it blows up to infinity.

\begin{figure}[h]
\centering%
\begin{subfigure}[b]{0.49\textwidth}
                
                \subcaption{ $\lambda=0.5$}\label{fig_local_kappa_a}
                \includegraphics[width=\textwidth, keepaspectratio=true]{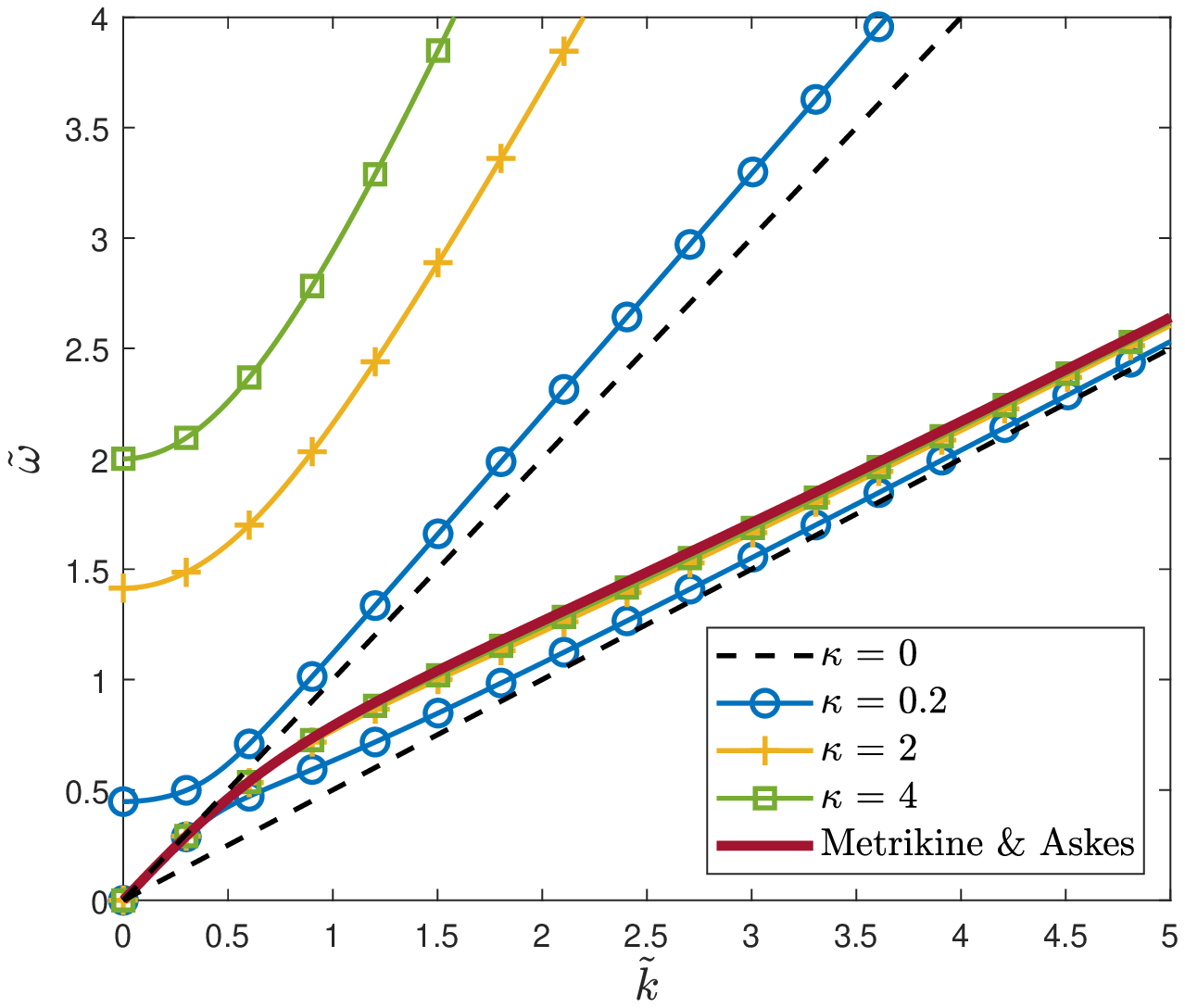}
\end{subfigure}
\begin{subfigure}[b]{0.49\textwidth}
                 
                \subcaption{ $\lambda=2$}\label{fig_local_kappa_b}
                \includegraphics[width=\textwidth, keepaspectratio=true]{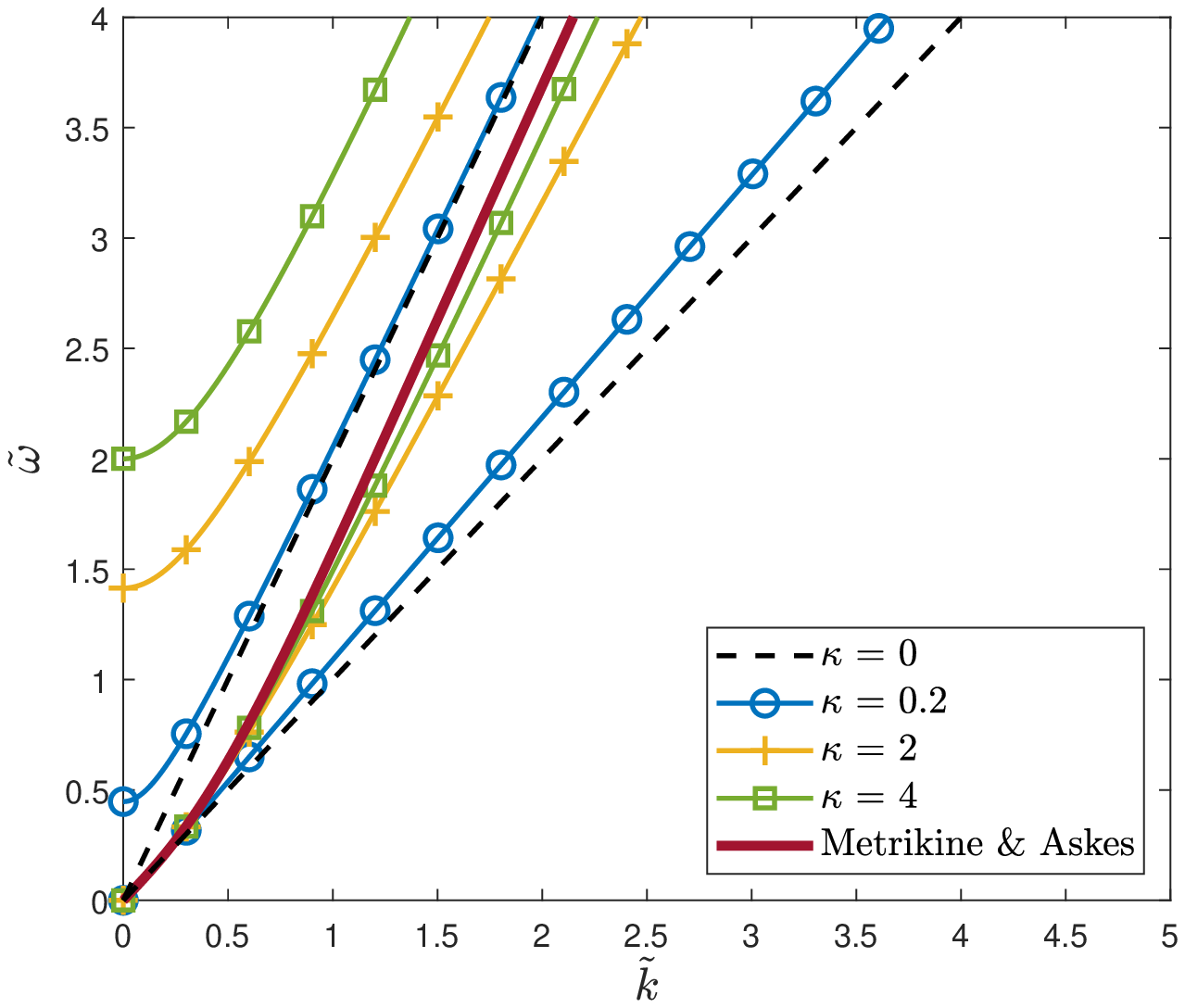}
\end{subfigure}
                \caption{The effect of parameter $\kappa$ on the dimensionless dispersion curve for the local micromorphic model.
                }
                \label{fig_local_kappa}
\end{figure} 

Note that the dispersion curves in Figure~\ref{fig_local_kappa_a} are plotted for a fixed value of $\lambda=0.5<1$ and in Figure~\ref{fig_local_kappa_b} for a fixed value of $\lambda=2>1$. In the former case, $\lambda<\sqrt{1+\kappa}$ for
all meaningful (non-negative) values of $\kappa$, and so all the acoustic
branches have the same terminal slope, given by $\lambda$,
while the optical branches have terminal slopes that
increase with $\kappa$ and are given by $\sqrt{1+\kappa}$;
see Figure~\ref{fig_local_kappa_a}.
In the latter case (i.e., $\lambda=2$), the acoustic
branches for $\kappa<3$ have terminal slope $\sqrt{1+\kappa}$, increasing with $\kappa$, and for all $\kappa\ge 3$ they have
the same terminal slope 2; see Figure~\ref{fig_local_kappa_b}. The optical branches
have the same terminal slope 2 for $\kappa\le 3$
and their slope increases with increasing $\kappa$
for $\kappa>3$. In all cases discussed here, the 
acoustic branch keeps rising to infinity as $\tilde k\to\infty$, and so there can be no band gap.

The influence of parameter $\lambda$ is illustrated in Figure \ref{fig_local_lambda} for a fixed value of $\kappa=4$. All the optical branches start from
the same point at frequency $\tilde\omega_0=\sqrt{\kappa}=2$. 
If $\lambda$ increases but remains below $\sqrt{1+\kappa}=\sqrt{5}$, the terminal slope of the acoustic branch is given by $\lambda$ and thus increases, while
the terminal slope of the optical branch remains constant, equal to $\sqrt{5}$. For higher values of $\kappa$, the terminal
slope of all acoustic branches is the same and equal to $\sqrt{5}$ while the terminal slope of the optical branch increases with increasing $\lambda$. 

\begin{figure}[h]
    \centering
    \includegraphics[width=0.5\textwidth, keepaspectratio=true]{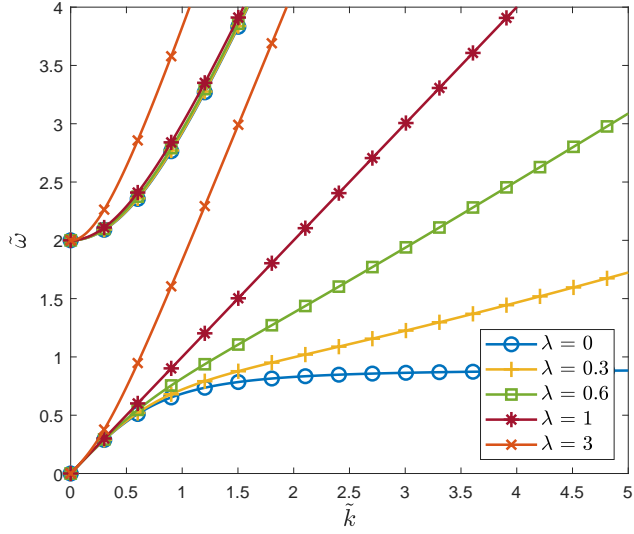}
    \caption{The effect of parameter $\lambda$ on the dimensionless dispersion curve for the local micromorphic model
    (parameter $\kappa=4$ is fixed).}
    \label{fig_local_lambda}
\end{figure}

Interesting behavior can be observed when  $\lambda=0$. This is the only case in which the acoustic branch has a finite limit $\sqrt{\kappa/(1+\kappa)}$ for large wave numbers $\tilde k$. Since the optical
branch always starts at $\tilde\omega_0=\sqrt{\kappa}$
and the frequency on this branch increases with increasing wave number, 
a band gap in the dispersion diagram appears; see the 
blue curves with circular markers in Figure \ref{fig_local_lambda}. 
This case corresponds to a vanishing micromorphic modulus $A$ and therefore to a model with micromorphic density
only, not accompanied by any micromorphic stiffness.

Another special case arises when $\lambda=1$, because then
the acoustic branch becomes a straight line of slope 1,
which happens to be the same as for standard local elasticity,
but there exists also an optical branch, given by $\tilde\omega=\sqrt{\kappa+(1+\kappa)\tilde k^2}$.
 For $\lambda<1$, the acoustic branch
is concave, and for $\lambda>1$ it is convex.
The optical branch is always convex.

\subsection{Integral micromorphic model}

In the previous subsection it was demonstrated that, for the local micromorphic model, a band gap in the one-dimensional dispersion curve appears only when the micromorphic modulus $A$ vanishes (and the micromorphic density parameter, $\eta$, is positive). In the following, the general integral micromorphic model introduced in Section~\ref{sec:2} is considered and the added value of nonlocal averaging is analyzed. 

Three typical nonlocal weight functions are considered here; their definitions and the corresponding Fourier images are summarized in Table \ref{tab:weightfun} and graphically
represented in Figure~\ref{fig_WF} for chosen length scale parameter $a=\ell$ (i.e., $\tilde a = 1$). The Gaussian weight function as well as the exponential one have an unbounded support and a strictly positive
Fourier image, while the quartic polynomial weight function vanishes for $\tilde{x}>\tilde{a}$ and its Fourier image
has negative values in certain intervals
of the wave number, the first one being
$\tilde{k}\in (5.763,9.095)$.

\begin{table}[]
\centering
\begin{tabular}{lll} \hline \\[-3mm]
 weight function & ${\tilde \alpha(\tilde k)}$ & ${\tilde \alpha^{*}}(\tilde k)$ \\[1mm] \hline \\[-2mm]
Gaussian  & $\dfrac{1}{\tilde a\sqrt{\pi}}{\rm e}^{-\tilde{x}^2/\tilde a^2}$ & ${\rm e}^{-\tilde{k}^2 \tilde a^2/4}$ \\[3mm]
exponential  & $\dfrac{1}{2 \tilde a}{\rm e}^{-|\tilde x|/\tilde a}$  & $\dfrac{1}{1+\tilde a^2\tilde k^2}$ \\[3mm]
quartic polynomial  & $ \dfrac{15}{16\tilde a}\ \left< 1 - \tilde x^2 /\tilde a^2\right>^2$ & $\dfrac{15}{\tilde k^5 \tilde a^5} \left(( 3-\tilde k^2\tilde a^2 ) \sin{(\tilde k \tilde a)} - 3\tilde k \tilde a \cos{(\tilde k \tilde a)} \right)$
\end{tabular}
\caption{Nonlocal weight functions and their Fourier images.}
\label{tab:weightfun}
\end{table}

\begin{figure}[h]
\centering%
\begin{subfigure}[b]{0.49\textwidth}
                
                \subcaption{Weight functions}\label{fig_kernel}               \includegraphics[width=\textwidth, keepaspectratio=true]{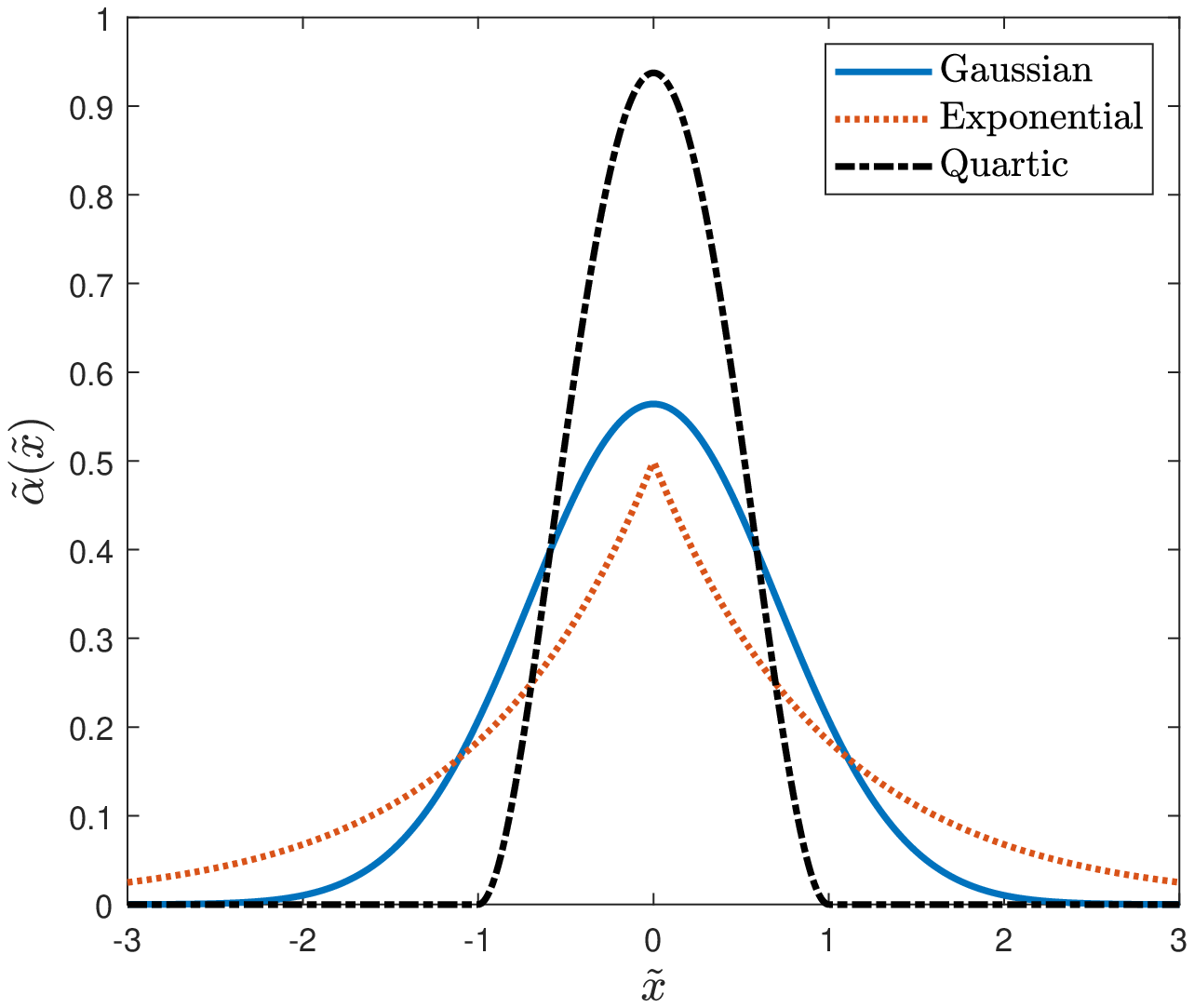}
\end{subfigure}
\begin{subfigure}[b]{0.49\textwidth}
               
                \subcaption{Fourier transforms of the weight functions}\label{fig_kernelF} \includegraphics[width=\textwidth, keepaspectratio=true]{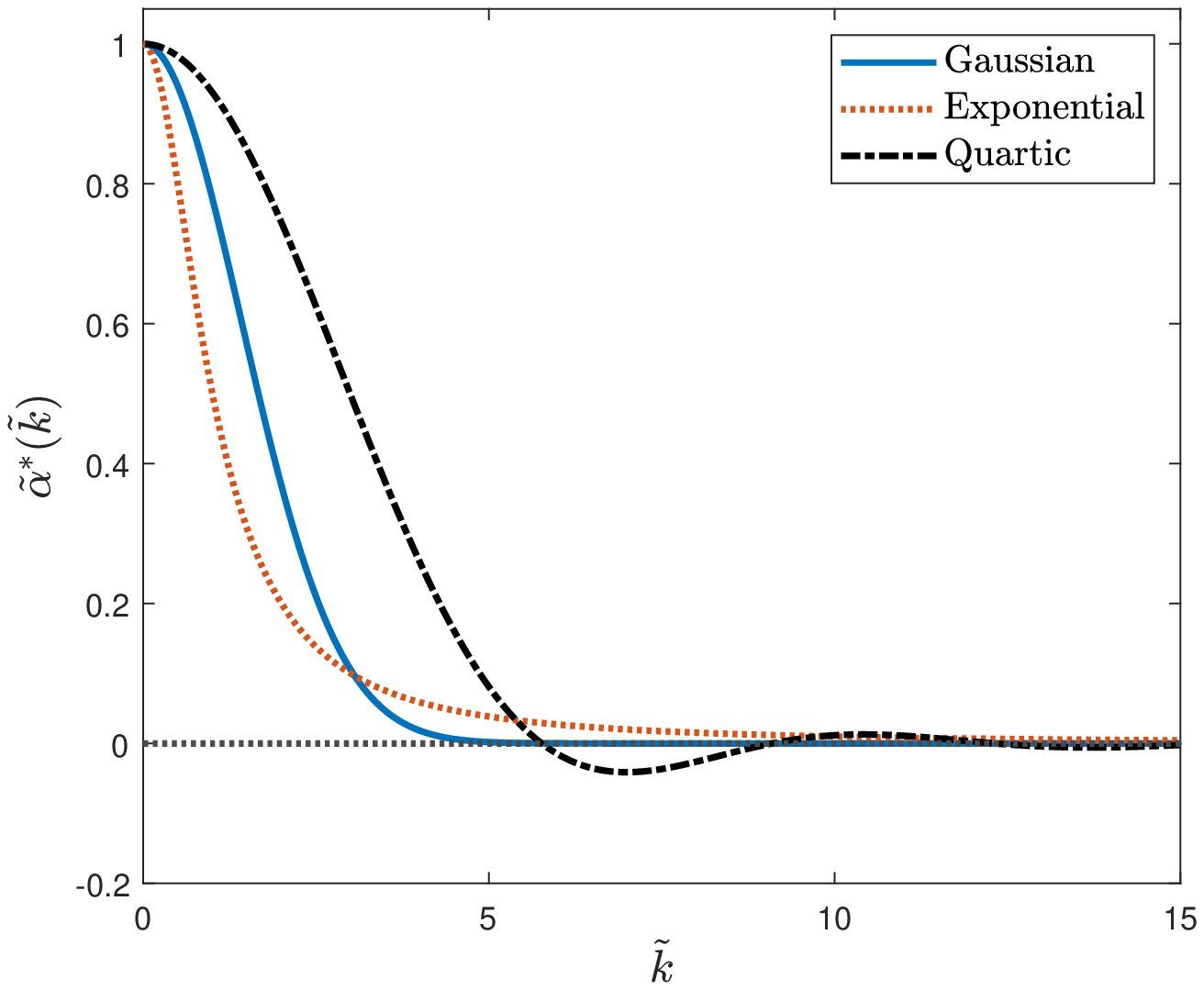} 
\end{subfigure}

                \caption{Graphs of the considered weight functions and their Fourier images.
                }
                \label{fig_WF}
\end{figure} 

Let us firstly investigate the asymptotic behavior of the dispersion relation when $\tilde k\to\infty$ (i.e., in the short-wave limit). For all nonlocal kernels listed in Table \ref{tab:weightfun}, the product $\tilde k{\tilde \alpha^{*}}(\tilde k)$ tends to zero as $\tilde k\to\infty$, and so equation\ (\ref{eq:dispnlmm3}) can be, in the short-wave range, approximated by
\beq\label{eq_disp_ap}
2\tilde\omega^2 
\approx  
\kappa+ \tilde{k}^2 \lambda^2+\tilde{k}^2 \tilde\alpha_0^*(\tilde k) \pm\sqrt{(\kappa+ \tilde{k}^2\lambda^2- \tilde{k}^2\tilde\alpha_0^*(\tilde k))^2}=\kappa+ \tilde{k}^2 \lambda^2+\tilde{k}^2 \tilde\alpha_0^*(\tilde k) \pm \lvert {\kappa+ \tilde{k}^2\lambda^2- \tilde{k}^2\tilde\alpha_0^*(\tilde k)\rvert }
\eeq     
For sufficiently large wave numbers, the term inside the absolute value brackets is typically positive, i.e., we have $\kappa+ \tilde{k}^2\lambda^2> \tilde{k}^2\tilde\alpha_0^*(\tilde k)$. The optical branch, for which the sign in front of the absolute value is positive, is then for large wave numbers approximated by the relation $\tilde\omega\approx \sqrt{\kappa + \tilde k^2\lambda^2}$. For $\lambda\neq 0$, this further reduces to $\tilde\omega\approx  \lambda\tilde k$, meaning that the branch asymptotically approaches a straight line of slope $\lambda=\sqrt{A\rho/(E\eta)}$. On the other hand, the  acoustic branch is for short waves approximated by
\beq\label{eq:acoust}
\tilde\omega\approx\tilde k\sqrt{{\tilde \alpha_0^{*}}(\tilde k)}
\eeq  
Therefore, the behavior of the acoustic branch for large wave numbers depends on the properties of the chosen kernel. 

For the Gaussian weight function, the right-hand side of equation\ \eqref{eq:acoust} tends to zero as $\tilde k \to \infty$. For the exponential kernel, the right-hand side of equation\ \eqref{eq:acoust} has limit $1/\tilde a$, hence the acoustic branch approaches a horizontal line given by $\tilde\omega = 1/\tilde a$. 
In the case of the quartic weight function, the Fourier image ${\tilde \alpha_0^{*}}(\tilde k)$ is an oscillatory function and hence the frequencies are imaginary for wave numbers 
in those intervals in which the Fourier image is negative. 

Since all weight functions considered here satisfy the condition that $\alpha_0^*(\tilde k)\to 0$ as $\tilde{k}\to\infty$,
the aforementioned assumption that $\kappa+ \tilde{k}^2\lambda^2> \tilde{k}^2\tilde\alpha_0^*(\tilde k)$
for sufficiently large wave numbers $\tilde{k}$
could be violated only if $\lambda=0$ and the 
exponential weight function $\alpha_0$ is used.
For this weight function, the product 
$\tilde{k}^2\tilde\alpha_0^*(\tilde k)$ tends to $1/\tilde{a}^2$ from below, and the assumption
is violated if $\kappa\tilde{a}^2\le 1$.
In this special case, the optical branch would approach a horizontal straight line given by
$\tilde{\omega}=1/\tilde{a}$, and the acoustic branch would approach a horizontal straight
line given by $\tilde{\omega}=\sqrt{\kappa}$. 
Recall that $\lambda=0$ corresponds to the model
with $A=0$, i.e., with no micromorphic stiffness.

The influence of nonlocal averaging is visualized in Figure \ref{fig_nonlocal_1}
for two combinations of parameters $\kappa$ and $\lambda$ and for the three weight functions described in Table \ref{tab:weightfun}. In order to reduce the number of independent model parameters, both kernels ($\tilde \alpha_0$ and $\tilde \alpha_1$) 
are assumed to be identical, including their
length parameters $\tilde a_0=\tilde a_1$. Note that the approximations considered in  \eqref{eq_disp_ap} hold only for the nonlocal case, since the product $\tilde k\tilde \alpha_1^{*}(\tilde k)$
tends to infinity for $\tilde k\to\infty$ when the averaging kernel is considered as the Dirac delta distribution, i.e., in the local case. Therefore, the blue curves in Figure \ref{fig_nonlocal_1}, plotted for $\tilde a_0=\tilde a_1=0$ (local case), differ in their asymptotic
behavior for large wave numbers from all the other cases with $\tilde a_0=\tilde a_1>0$.  
They correspond to the local micromorphic model
and are, in Figure \ref{fig_nonlocal_1a}, \ref{fig_nonlocal_1c}, and \ref{fig_nonlocal_1e}, identical with the green curves
in Figure~\ref{fig_local_kappa_a}, and in Figure  \ref{fig_nonlocal_1b}, \ref{fig_nonlocal_1d}, and \ref{fig_nonlocal_1f} identical with the orange curves
in Figure~\ref{fig_local_kappa_b}.

For small wave numbers, the dispersion curves
of the nonlocal model remain close to those
of the local model, because the effects of
nonlocal averaging are negligible if the 
wave length is much larger than the characteristic length of the averaging
operator. For higher wave numbers, the
nonlocal curves deviate from the local ones,
and this deviation occurs earlier for larger
values of the length scale parameter $\tilde a_0=\tilde a_1$.

The behavior of the {\bf optical branch} is qualitatively similar for all the investigated weight functions, but differs for the two combinations of parameters $\kappa$ and $\lambda$ selected as representative examples. For $\kappa=4$ and $\lambda=0.5$, the optical branch
attains a local maximum and in a certain range descends, but then rises again and approaches
an inclined straight asymptote, as predicted above;
see Figure \ref{fig_nonlocal_1a}, \ref{fig_nonlocal_1c}, and \ref{fig_nonlocal_1e}. On the other hand,
for $\kappa=2$ and $\lambda=2$, the optical branch 
rises monotonically, at least in the range plotted
in Figure \ref{fig_nonlocal_1b}, \ref{fig_nonlocal_1d}, and \ref{fig_nonlocal_1f}. In both cases,
the global minimum on the optical branch occurs
 at $\tilde k=0$. 
 
 As mentioned earlier, the behavior of the {\bf acoustic branch} depends on the considered kernel function. For the Gaussian function, the acoustic branch attains a maximum and then
descends and asymptotically approaches the
horizontal axis. For the exponential kernel, the acoustic branch has a finite limit equal to $1/\tilde a_0$. Even though it may seem from Figure \ref{fig_nonlocal_1c} and \ref{fig_nonlocal_1d}  that the acoustic branch approaches the limit value from below, it is not the case and the maximum of the branch is reached first and then the curve  approaches the horizontal asymptote from above. For the quartic weight function, the acoustic branch first reaches the maximum and then starts to descend. As discussed before, for larger wave numbers the curve may be approximated by equation\ \eqref{eq:acoust}. However, since the Fourier image $\tilde \alpha_0^{*}(\tilde{k})$ oscillates and changes sign for the quartic function, in some intervals the frequencies become imaginary and thus cannot be visualized in Figures \ref{fig_nonlocal_1e} and \ref{fig_nonlocal_1f}.

\begin{figure}[H]
\centering%
\begin{subfigure}[b]{0.48\textwidth}
               \subcaption{ $\kappa=4$, $\lambda=0.5$, \textbf{Gaussian} kernel}\label{fig_nonlocal_1a} \includegraphics[width=0.9\textwidth, keepaspectratio=true]{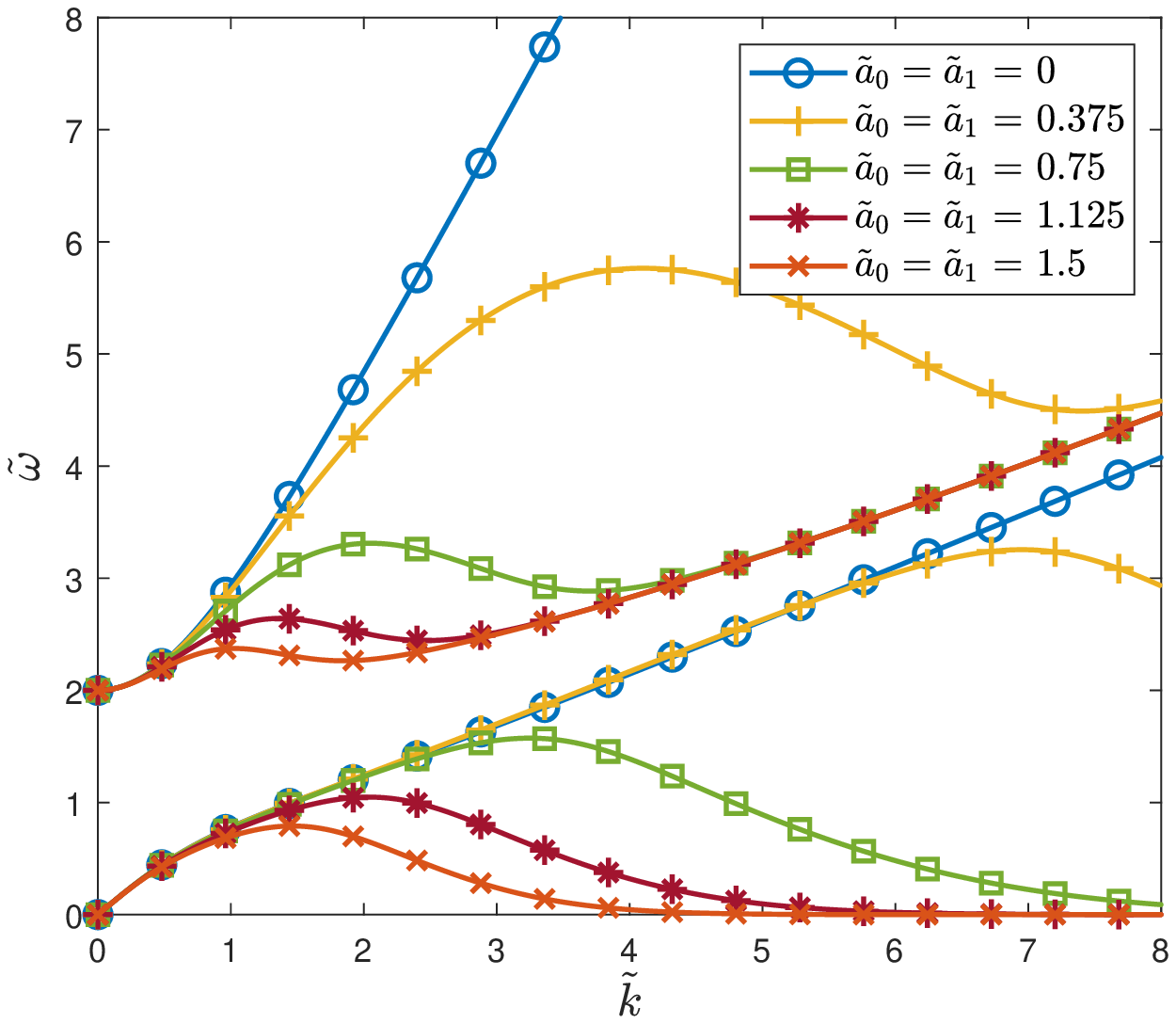}
\end{subfigure}
\begin{subfigure}[b]{0.48\textwidth}
               \subcaption{ $\kappa=2$, $\lambda=2$, \textbf{Gaussian} kernel}\label{fig_nonlocal_1b} \includegraphics[width=0.9\textwidth, keepaspectratio=true]{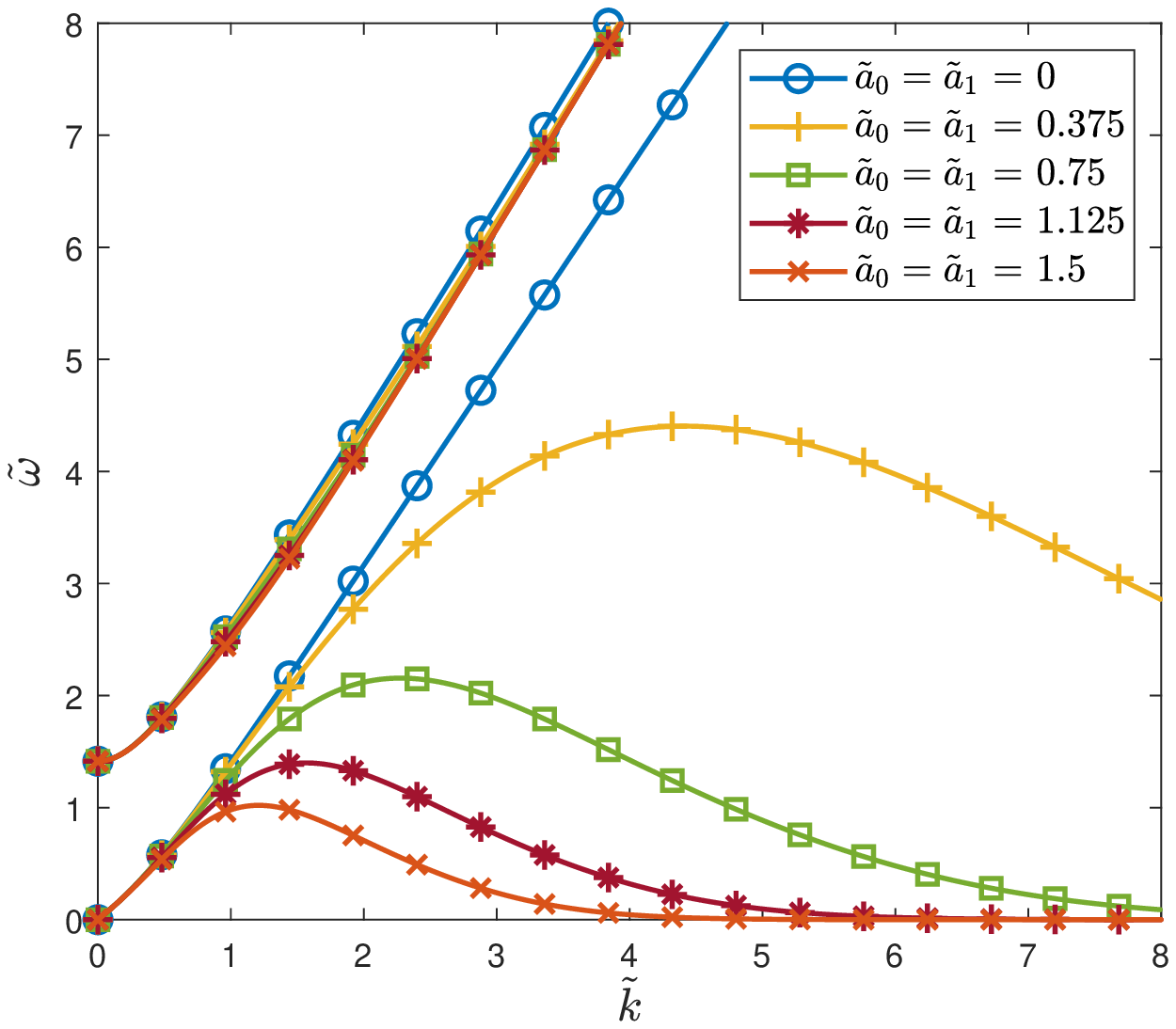} 
\end{subfigure}
\begin{subfigure}[b]{0.48\textwidth}
                
                \subcaption{ $\kappa=4$, $\lambda=0.5$, \textbf{exponential} kernel}  \label{fig_nonlocal_1c}
             \includegraphics[width=0.9\textwidth, keepaspectratio=true]{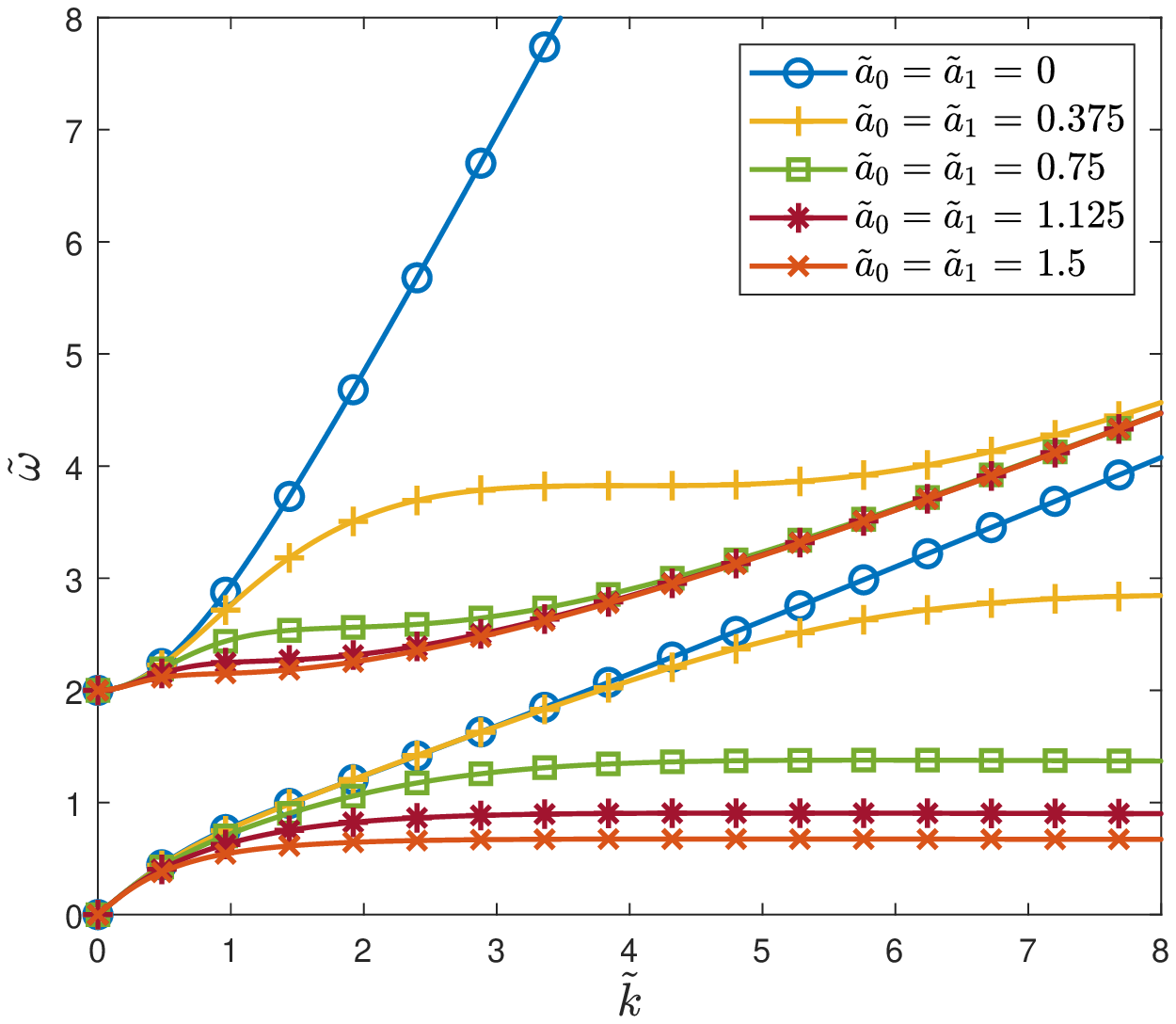}
\end{subfigure}
\begin{subfigure}[b]{0.48\textwidth}
                 
                \subcaption{ $\kappa=2$, $\lambda=2$, \textbf{exponential} kernel}\label{fig_nonlocal_1d}
                \includegraphics[width=0.9\textwidth, keepaspectratio=true]{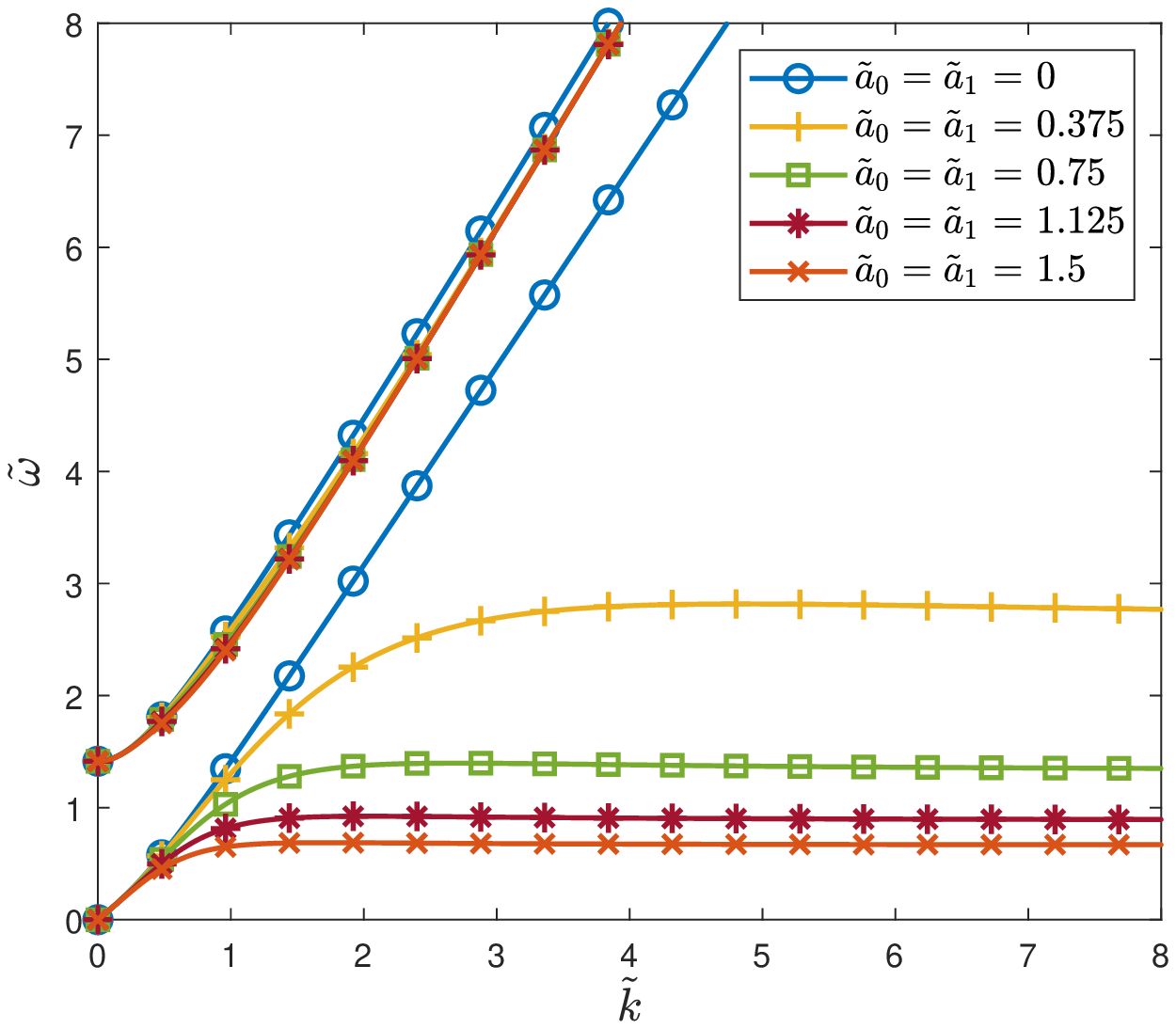}
\end{subfigure}
\begin{subfigure}[b]{0.48\textwidth}
                
                \subcaption{ $\kappa=4$, $\lambda=0.5$, \textbf{quartic} kernel}  \label{fig_nonlocal_1e}
                \includegraphics[width=0.9\textwidth, keepaspectratio=true]{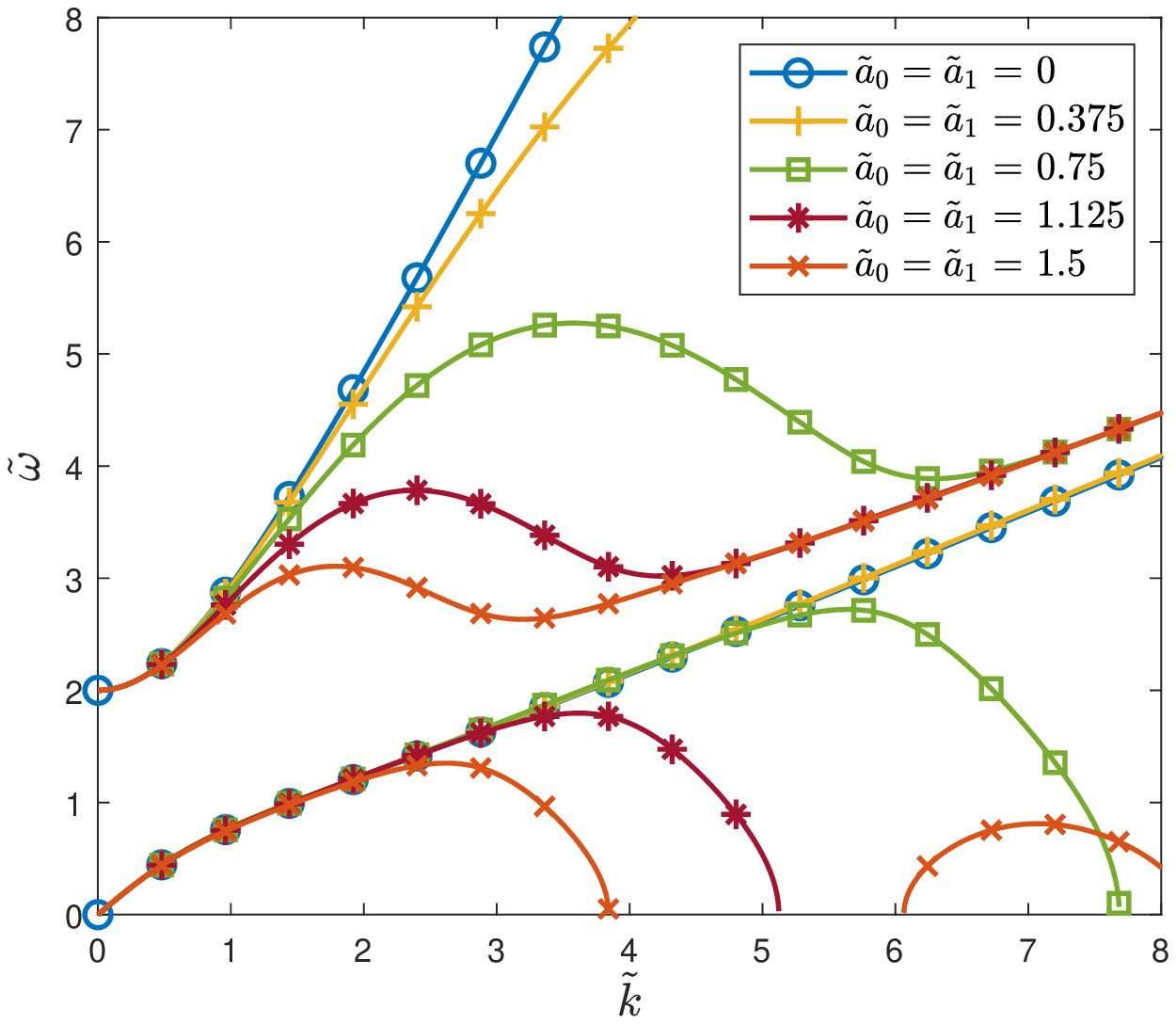}
\end{subfigure}
\begin{subfigure}[b]{0.48\textwidth}
                 
                \subcaption{ $\kappa=2$, $\lambda=2$, \textbf{quartic} kernel}\label{fig_nonlocal_1f}
                \includegraphics[width=0.9\textwidth, keepaspectratio=true]{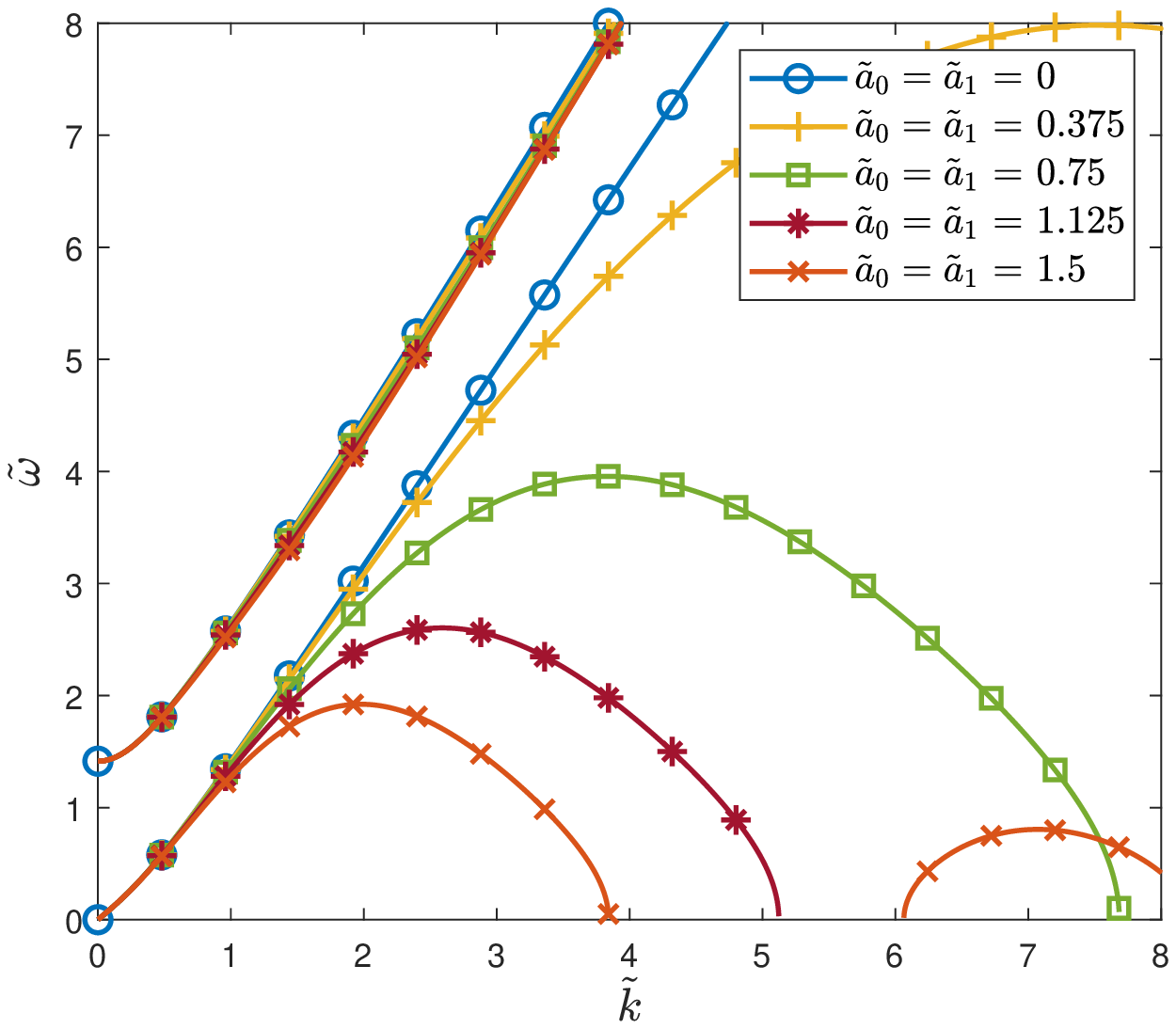}
\end{subfigure}
                \caption{The effect of the nonlocal averaging on the dispersion curve of the integral micromorphic model visualized for various characteristic lengths $\tilde a_0=\tilde a_1$.
                }
                \label{fig_nonlocal_1}
\end{figure}

Note that the smaller value on the right-hand side of (\ref{eq:dispnlmm3}) becomes negative
if 
\beq 
(\kappa+\tilde{k}^2\lambda^2)\tilde{\alpha}_0^*(\tilde{k}) + \tilde{k}^2\lambda^2\kappa\tilde{\alpha}_1^{*2}(\tilde{k})<0
\eeq
For weight functions with strictly positive Fourier images, this can never happen, and so both frequencies computed from (\ref{eq:dispnlmm3}) are real, for an arbitrary wave number $\tilde{k}$. However, if weight function $\alpha_0$ is the quartic one, the condition can be satisfied for some wave numbers, and the frequency on the acoustic branch is then imaginary. In particular,
if both weight functions are the same,
$\alpha_0=\alpha_1$, then the imaginary 
frequencies appear for those wave numbers
for which 
\beq 
-\frac{1}{\kappa}-\frac{1}{\tilde{k}^2\lambda^2}<\tilde{\alpha}_0^*(\tilde{k})<0
\eeq
The lowest positive wave number for which the frequency becomes zero is the lowest one for
which the Fourier image becomes zero,
which is $\tilde{k}\approx 5.763/\tilde{a}$.
For instance, for $\tilde{a}_0=\tilde{a}_1=1.5$,
one obtains $\tilde{k}\approx 3.84$, which 
is in agreement with the first intersection of the 
light red curve with the horizontal axis
in Figure~\ref{fig_nonlocal_1e} and \ref{fig_nonlocal_1f}. Note that the 
position of this point is unaffected by
parameters $\kappa$ and $\lambda$. 
 
For sufficiently large $\tilde a_0$ and $\tilde a_1$, the maximum frequency on the acoustic branch is below the minimum frequency on the optical branch, and a {\bf band gap} is observed.
Figure \ref{fig_nonlocal_2} shows the mentioned frequency extrema for the two selected
combinations of parameters $\kappa$ and $\lambda$ and for the three considered weight functions. It is apparent that, for all averaging kernels, 
the minimum frequency on the optical branch is 
independent of parameters $\tilde a_0$ and $\tilde a_1$, and equals
$\sqrt{\kappa}$, while the maximum frequency on the acoustic branch decreases with increasing nonlocal
length scale. 
 The values of parameters $\tilde a_0=\tilde a_1$ above which a band gap starts to appear are displayed in Table \ref{tab:resul}. 
\begin{table}[]
\centering
\begin{tabular}{*3c}
weight function &  \multicolumn{2}{c}{$\tilde a_0=\tilde a_1$} \\
\hline
{} & $\kappa=4$, $\lambda=0.5$ & $\kappa=2$, $\lambda=2$ \\ 
\cline{2-3}
Gaussian & 0.595 & 1.114 \\
exponential & 0.525  & 0.741 \\
quartic & 1.012 & 1.994
\end{tabular}
\caption{Values of nonlocal parameters $\tilde a_0=\tilde a_1$ above which a band gap starts to appear.}
\label{tab:resul}
\end{table}

\begin{figure}[H]
\centering%
\begin{subfigure}[b]{0.49\textwidth}
                
                \subcaption{ $\kappa=4$, $\lambda=0.5$}\label{fig_nonlocal_2a}
                \includegraphics[width=\textwidth, keepaspectratio=true]{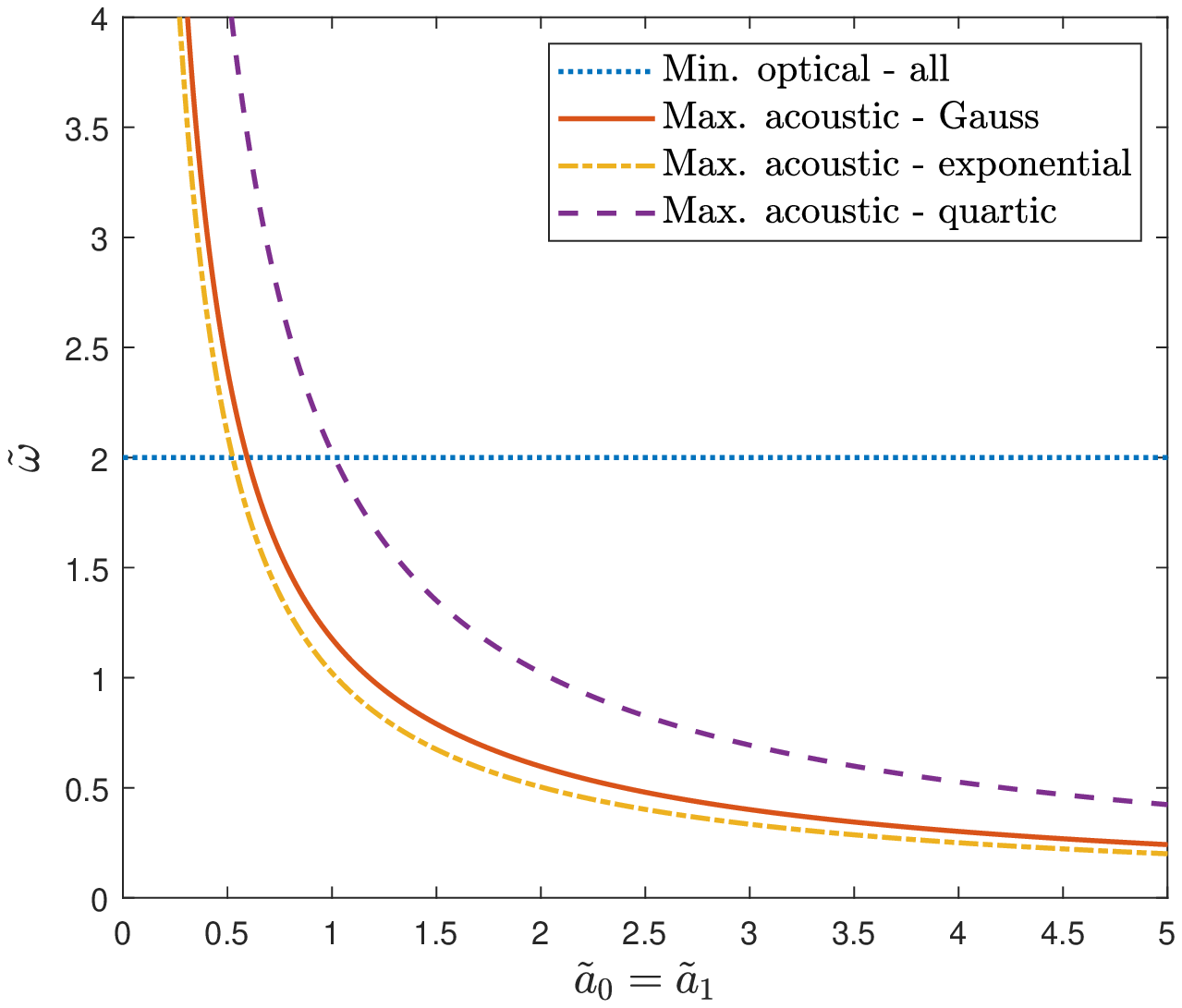}
\end{subfigure}
\begin{subfigure}[b]{0.49\textwidth}
                 
                \subcaption{ $\kappa=2$, $\lambda=2$}\label{fig_nonlocal_2b}
                \includegraphics[width=\textwidth, keepaspectratio=true]{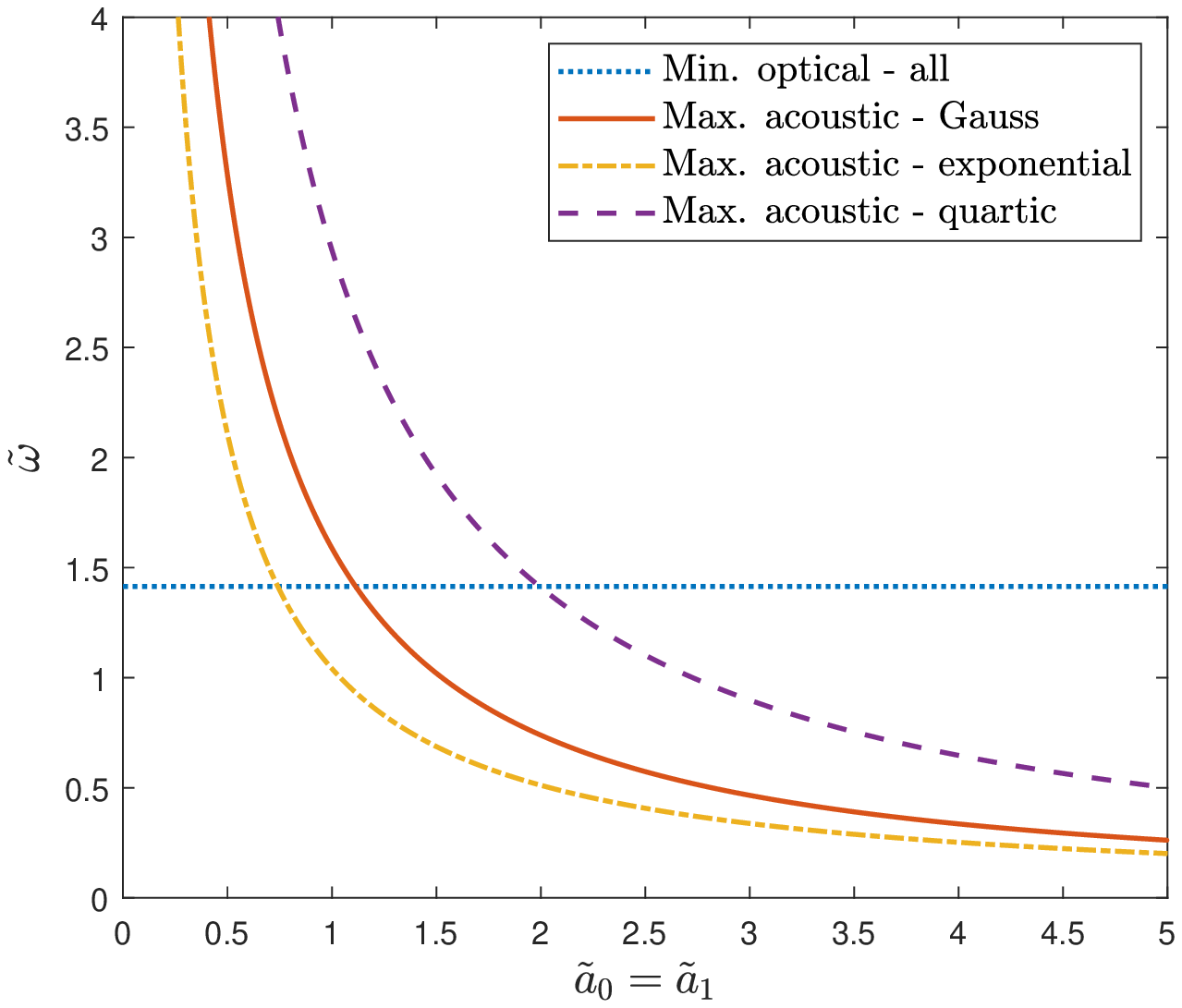}
\end{subfigure}

                \caption{Maximum frequency on the acoustic branch and minimum frequency on the optical branch as function of characteristic length $\tilde a_0=\tilde a_1$.
                }
                \label{fig_nonlocal_2}
\end{figure} 

So far, we have considered equal values
of length scale parameters $\tilde{a}_0$ and $\tilde a_1$.
Figure \ref{fig_nonlocal_3} shows what happens
if these parameters  are varied independently. 
The white regions correspond to parameter combinations for
which the dispersion diagram does not exhibit a band gap.
Other colors indicate the width of the band gap that 
forms for the respective combinations of $\tilde a_0$ and $\tilde a_1$.
In particular, it turns out that both parameters need
to have a certain minimum value for a band gap to exist; this holds for all the investigated weight functions and both selected combinations of parameters $\kappa$ and $\lambda$.
In other words, both nonlocal enrichments represented
in (\ref{eq_free_en}) by functions $\alpha_0$ and $\alpha_1$ are indispensable if the objective is
to construct an integral micromorphic
model that leads to a dispersion diagram with a band gap.
Free energy needs to depend on nonlocal strain, and 
the micromorphic variable needs to be linked to
(another or the same) nonlocal strain.
\begin{figure}[h]
\centering%
\begin{subfigure}[b]{0.48\textwidth}
               \subcaption{ $\kappa=4$, $\lambda=0.5$, \textbf{Gaussian} kernel}\label{fig_nonlocal_3a} \includegraphics[width=\textwidth, keepaspectratio=true]{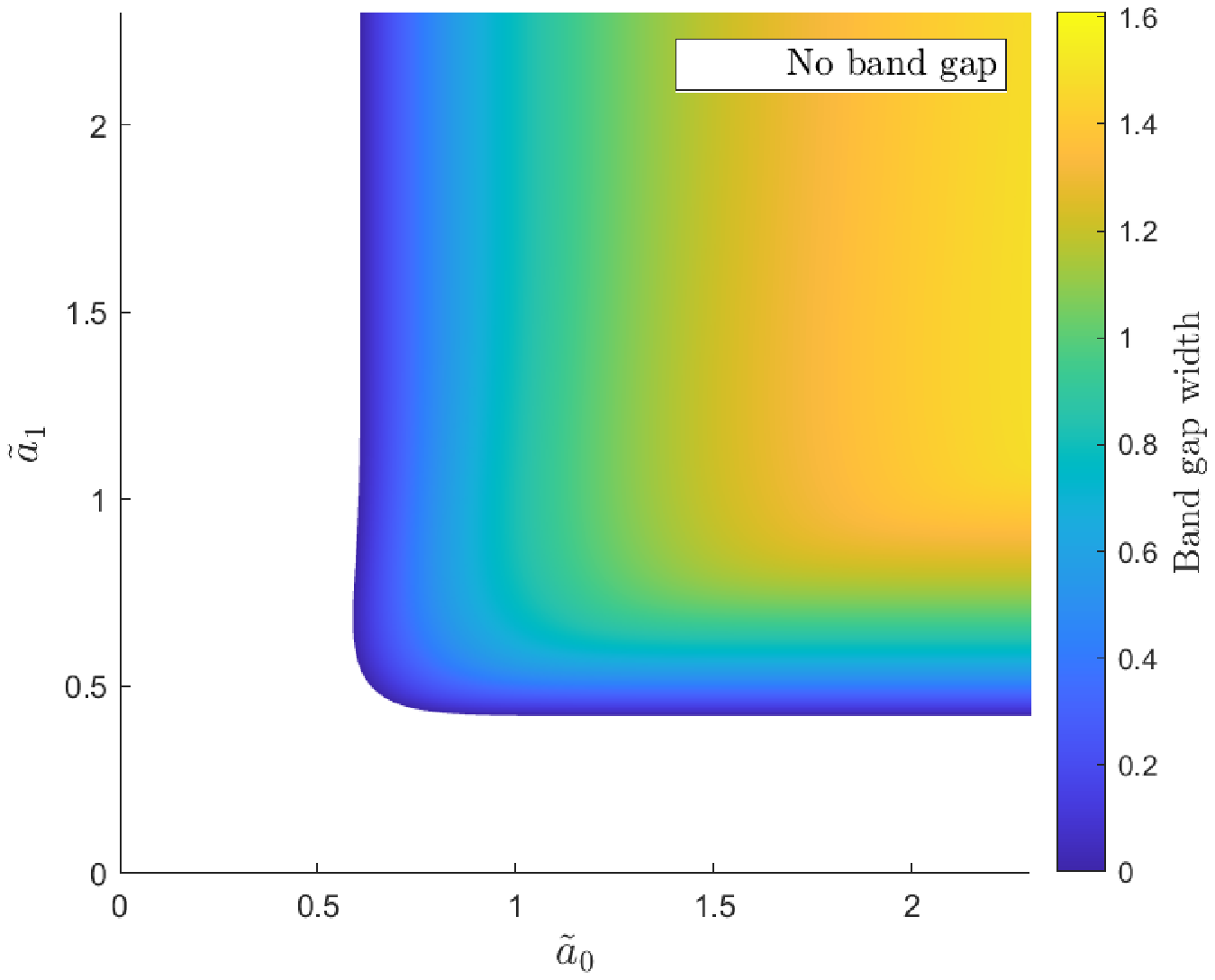}
                
\end{subfigure}
\begin{subfigure}[b]{0.48\textwidth}
                \subcaption{ $\kappa=2$, $\lambda=2$, \textbf{Gaussian} kernel} \label{fig_nonlocal_3b} \includegraphics[width=\textwidth, keepaspectratio=true]{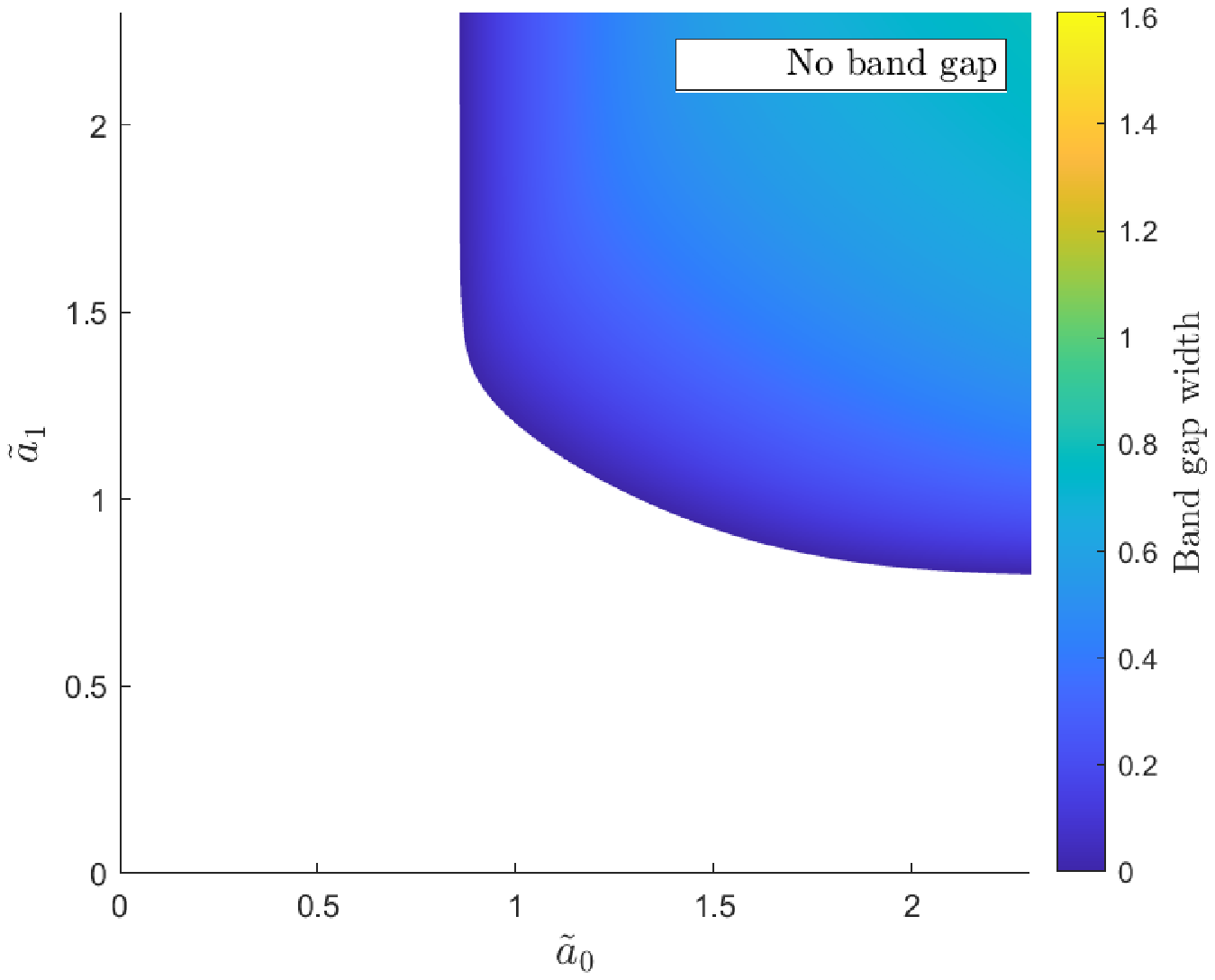} 
               
\end{subfigure}
\begin{subfigure}[b]{0.48\textwidth}
                \subcaption{ $\kappa=4$, $\lambda=0.5$, \textbf{exponential} kernel}\label{fig_nonlocal_3c} \includegraphics[width=\textwidth, keepaspectratio=true]{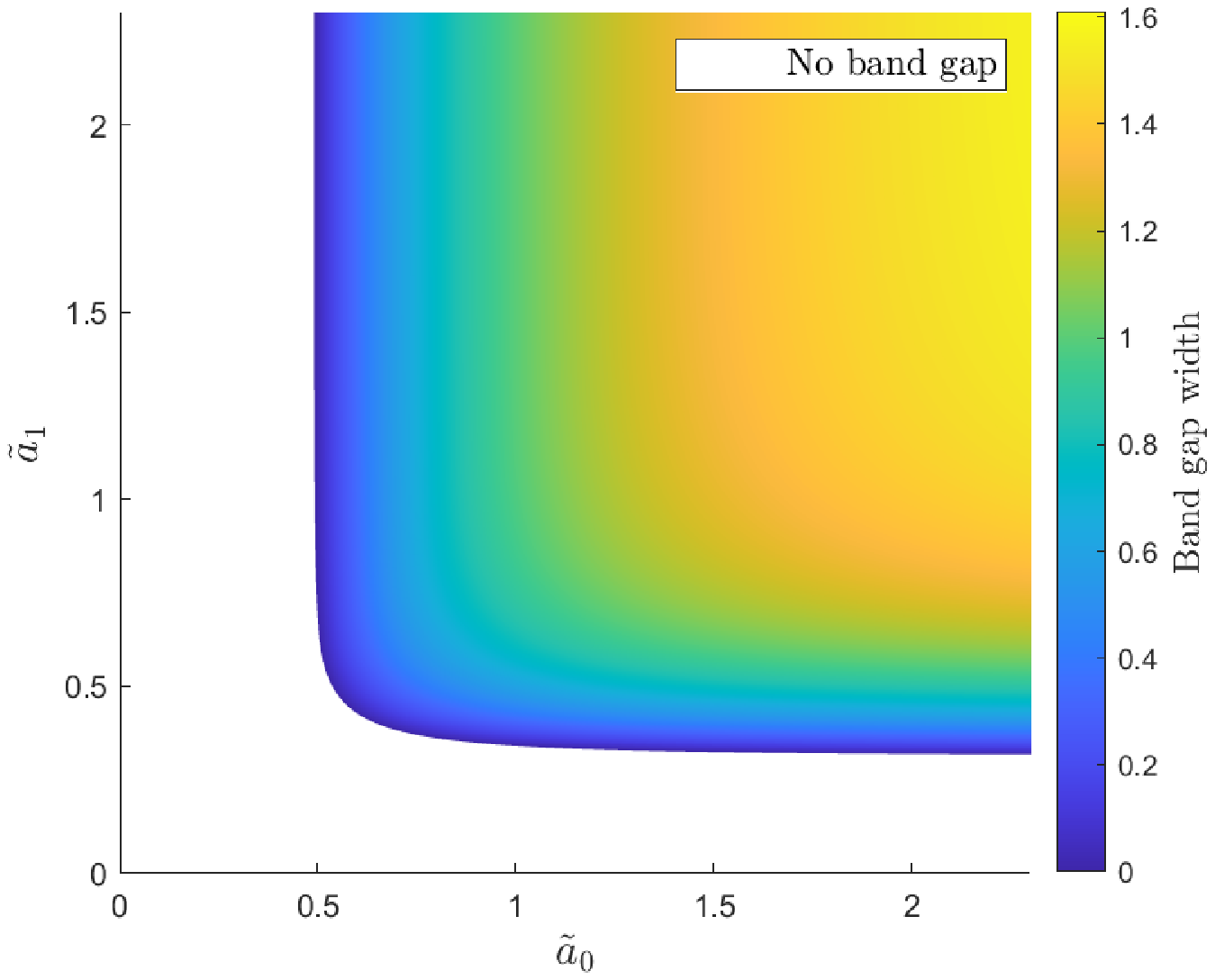}
               
\end{subfigure}
\begin{subfigure}[b]{0.48\textwidth}
                \subcaption{ $\kappa=2$, $\lambda=2$, \textbf{exponential} kernel} \label{fig_nonlocal_3d} \includegraphics[width=\textwidth, keepaspectratio=true]{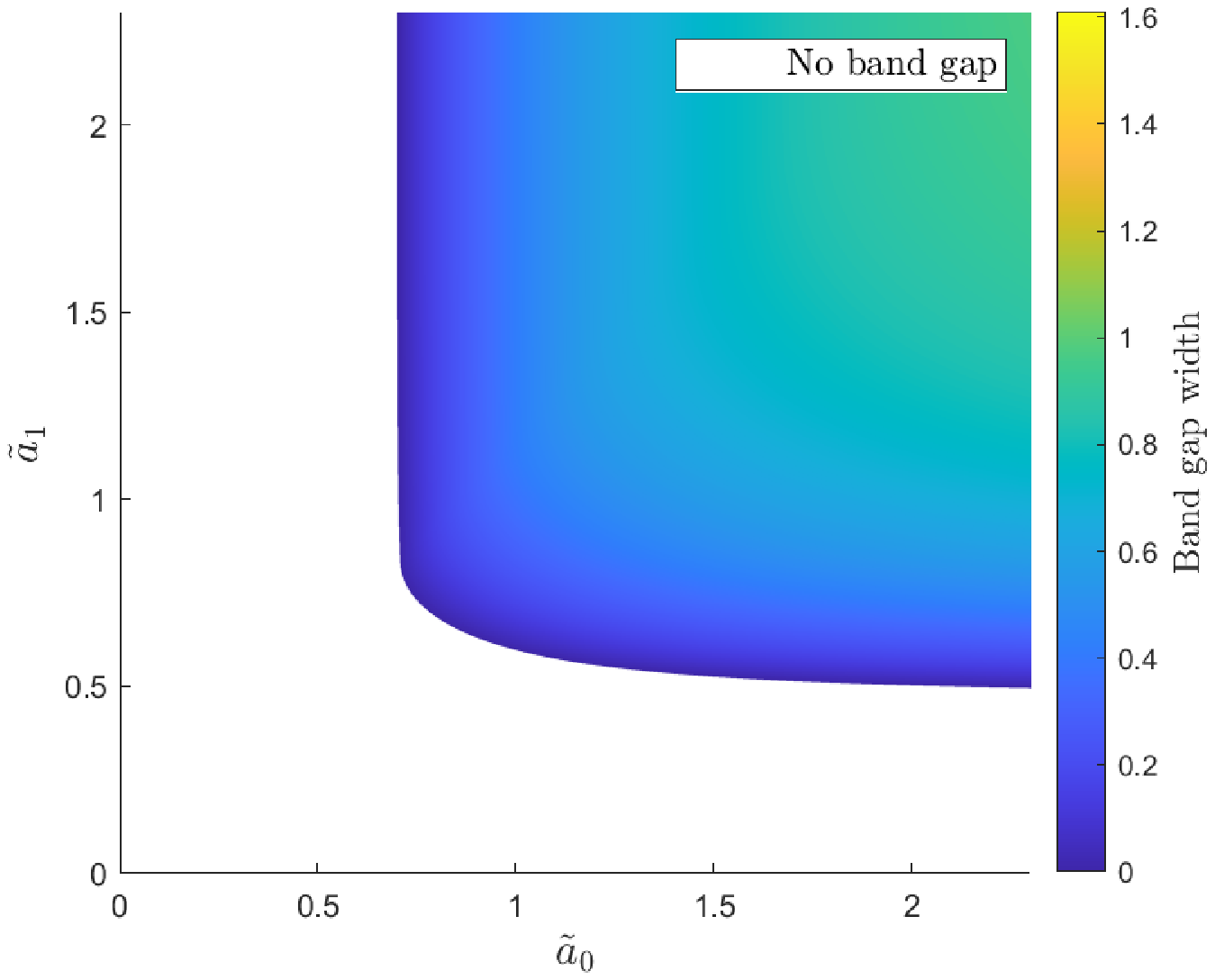} 
               
\end{subfigure}
\begin{subfigure}[b]{0.48\textwidth}
               \subcaption{ $\kappa=4$, $\lambda=0.5$, \textbf{quartic} kernel}\label{fig_nonlocal_3e} \includegraphics[width=\textwidth, keepaspectratio=true]{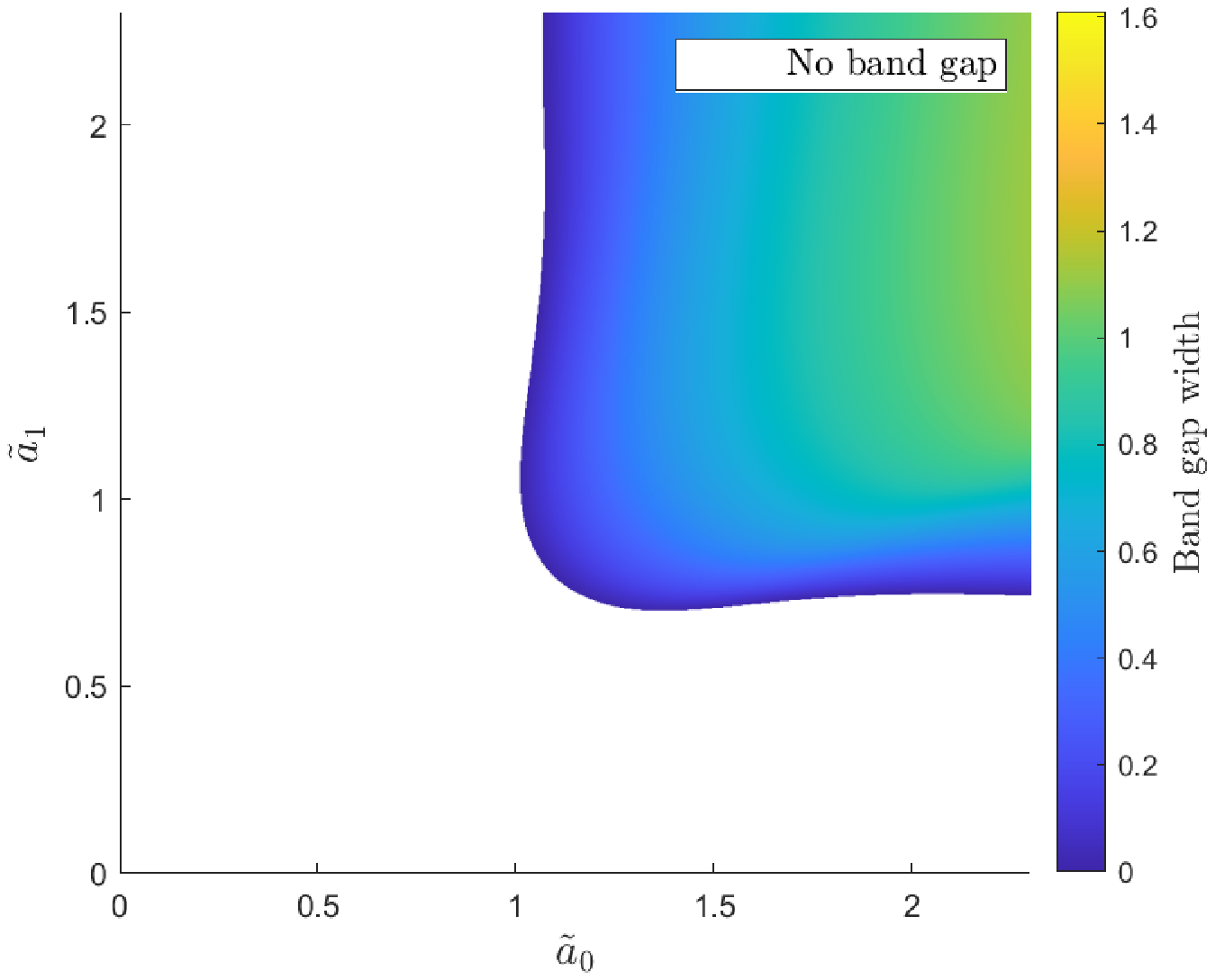}
                
\end{subfigure}
\begin{subfigure}[b]{0.48\textwidth}
               \subcaption{ $\kappa=2$, $\lambda=2$, \textbf{quartic} kernel} \label{fig_nonlocal_3f} \includegraphics[width=\textwidth, keepaspectratio=true]{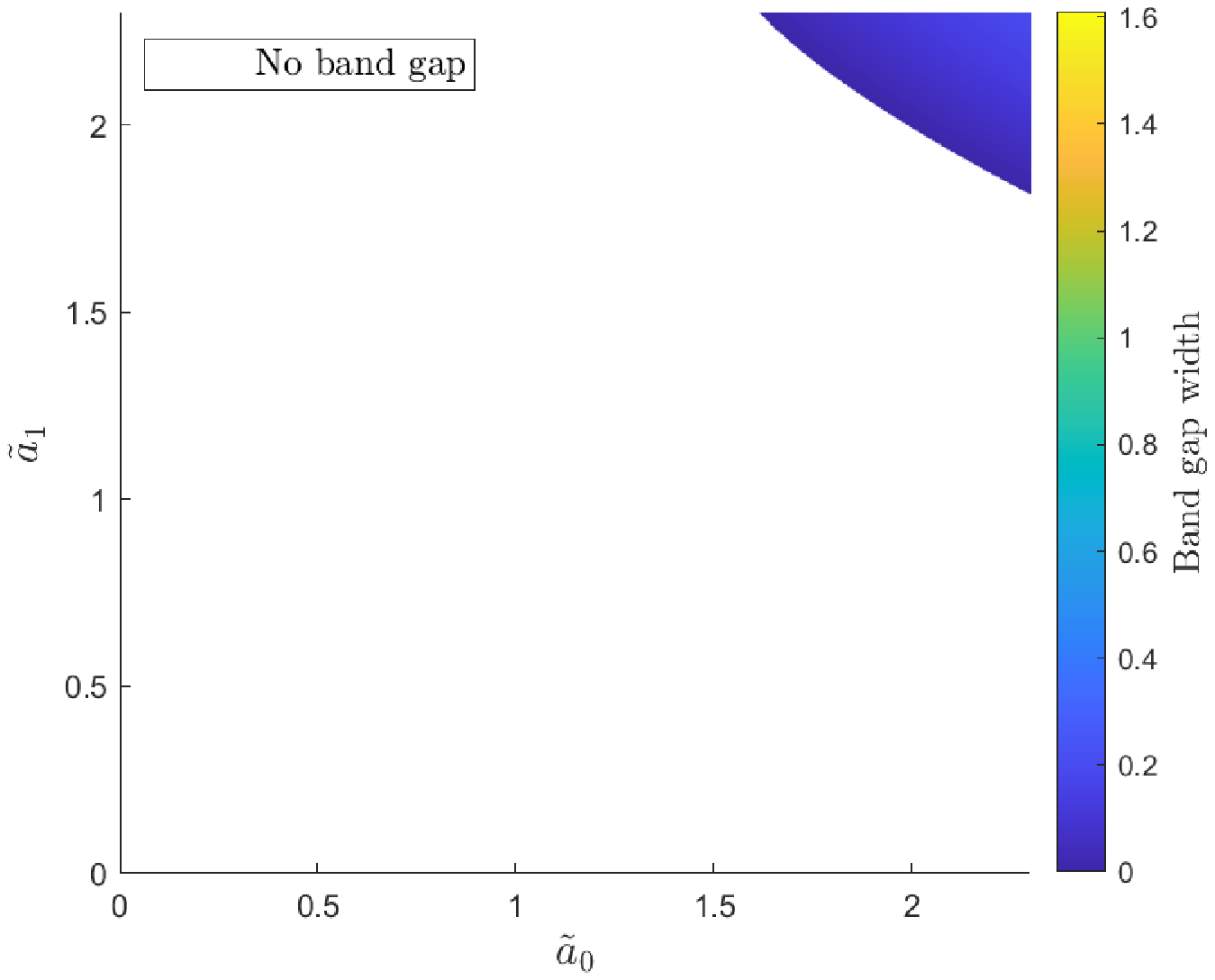} 
                
\end{subfigure}
                \caption{The width of the band gap depending on characteristic lengths $\tilde a_0$ and $\tilde a_1$ (white region corresponds to combinations of parameters for which no band gap exists).
                }
                \label{fig_nonlocal_3}
\end{figure} 

Finally, the width of the band gap  is studied for independently varying parameters $\kappa$ and $\lambda$ and for a fixed combination of internal length parameters.  In Figure \ref{fig_nonlocal_4}, the results are visualized for all three considered weight functions and for two
combinations of nonlocal parameters, $\tilde a_0=\tilde a_1=0.5$ and $\tilde a_0=\tilde a_1=0.7$.
It is easy to prove that for $\kappa=0$, one of the branches
of the dispersion diagram is linear and given by
$\tilde\omega = \lambda\tilde{k}$, and so no band gap
can exist for vanishing $\kappa$ (provided that $\lambda>0$). This is natural, since $\kappa=0$ means
$H=0$ and the micromorphic variable is then fully decoupled
from the displacement and strain. 

For nonlocal averaging
with the exponential kernel, the graphs in Figure~\ref{fig_nonlocal_4c} and \ref{fig_nonlocal_4d}
clearly show that a certain minimum
value of parameter $\kappa$ is needed for the band gap
to exist, and the most favorable case from this point of view occurs when $\lambda=0$. For this specific case,
the minimum required value of $\kappa$ can be determined analytically. Typical shapes of the dispersion diagram
for $\lambda=0$ and $\kappa=2$
are shown in Figure~\ref{fig10}. 
The blue curves with markers correspond to the local micromorphic model (special case
with $\tilde{a}_0=\tilde{a}_1=0$) and the other colors 
to the integral micromorphic model with various values of $\tilde{a}_0=\tilde{a}_1$. 
The minimum frequency on the optical
branch is attained for zero wave number and its value is
$\sqrt{\kappa}$, as already noted for the local model in Section~\ref{sec:3.1}. 
The acoustic branch is monotonically increasing from 0
and asymptotically approaches a finite limit value as
the wave number tends to infinity. For the local model,
it was mentioned in Section~\ref{sec:3.1} that the limit value 
is $\sqrt{\kappa/(1+\kappa)}$, which is always smaller than 
$\sqrt{\kappa}$, and so a band gap always exists but its width
is small for small values of $\kappa$. In contrast to that,
for the nonlocal model with exponential kernels, the right-hand
side of (\ref{eq:dispnlmm3}) with $\lambda=0$, $\tilde{\alpha}_0^*(\tilde{k})=1/(1+\tilde{a}_0^2\tilde{k}^2)$ and $\tilde{\alpha}_1^*(\tilde{k})=1/(1+\tilde{a}_1^2\tilde{k}^2)$ tends to 
$\kappa+1/\tilde{a}_0^2\pm |\kappa-1/\tilde{a}_0^2 |$ as $\tilde{k}\to\infty$. 
Therefore, if $\kappa\le 1/\tilde{a}_0^2$, the frequency on
the acoustic branch asymptotically tends to $\sqrt{\kappa}$
and no band gap exists. On the other hand, 
if $\kappa>1/\tilde{a}_0^2$, the frequency on
the acoustic branch asymptotically tends to $1/\tilde{a}_0$,
which is now smaller than $\sqrt{\kappa}$ and a 
 band gap is detected. We can thus conclude that,
 for the nonlocal model with an exponential kernel, 
 the band gap can exist only if parameter $\kappa$
 exceeds the minimum required value given by $1/\tilde{a}_0^2$.
 
 The foregoing analysis has been done for the exponential
 kernel.
 Interestingly,  for the other
two nonlocal kernels considered here (Gaussian and quartic),
a band gap can be found for arbitrarily small (positive)
values of $\kappa$, provided that $\lambda$ is also sufficiently small. 

If, for given values of parameters $\tilde a_0$ and $\tilde a_1$, parameter $\kappa$ is above its minimum value needed
for the formation of a band gap
(i.e., $1/\tilde{a}^2$ for the exponential kernel and 0
for the Gaussian or quartic kernel), the band gap exists
for parameter $\lambda$ ranging from 0 to a certain  upper limit, which in general
increases with increasing $\kappa$. Another general trend common to all studied weight functions is that, for increasing internal length parameters $\tilde a_0=\tilde a_1$, the domain in the $\lambda-\kappa$ space where band gap forms grows.    

\begin{figure}[h]
\centering%
\begin{subfigure}[b]{0.48\textwidth}
               \subcaption{ $\tilde a_0=\tilde a_1=0.5$, \textbf{Gaussian} kernel}\label{fig_nonlocal_4a} \includegraphics[width=\textwidth, keepaspectratio=true]{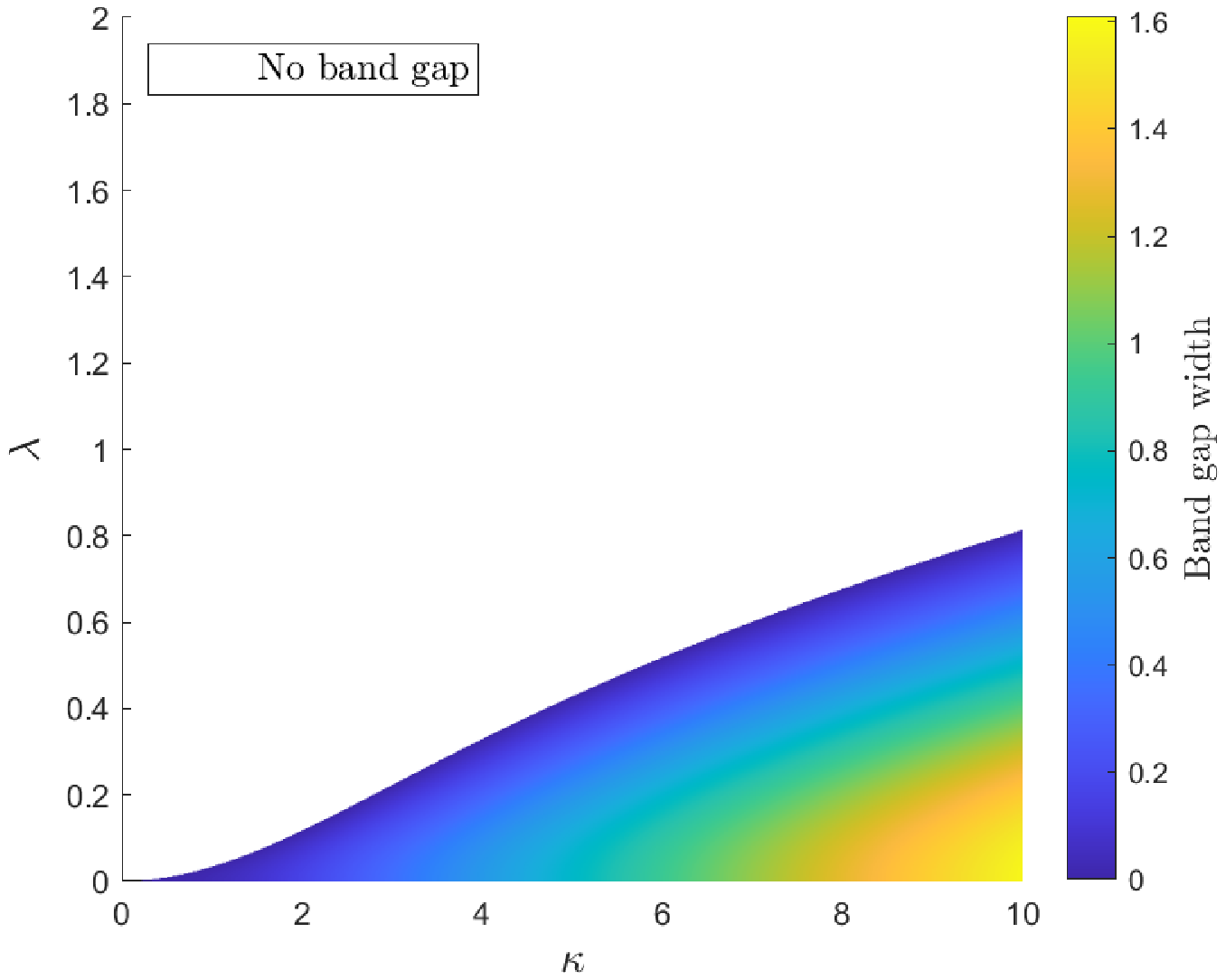}
                
\end{subfigure}
\begin{subfigure}[b]{0.48\textwidth}
               \subcaption{ $\tilde a_0=\tilde a_1=0.7$, \textbf{Gaussian} kernel} \label{fig_nonlocal_4b} \includegraphics[width=\textwidth, keepaspectratio=true]{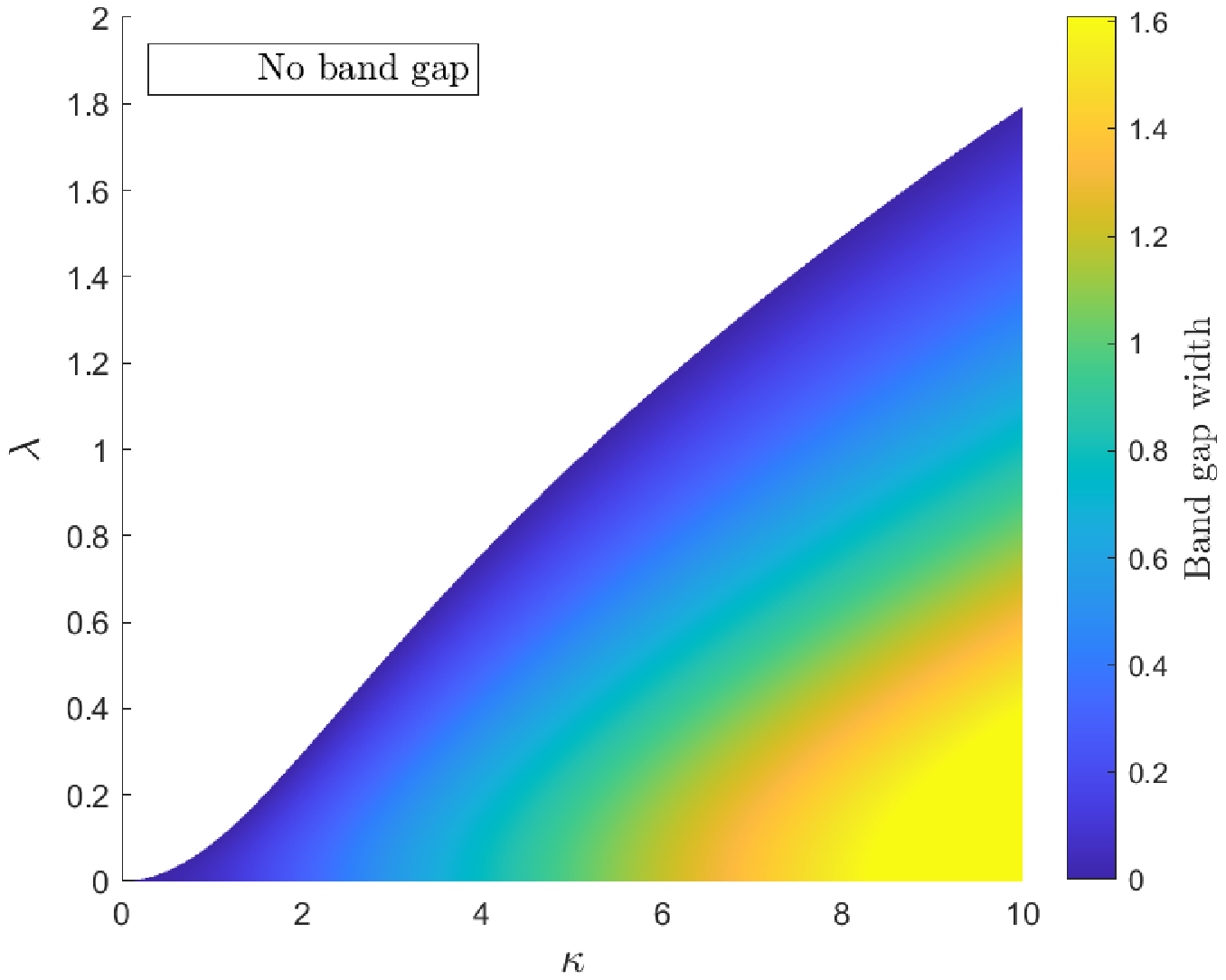} 
                
\end{subfigure}
\begin{subfigure}[b]{0.48\textwidth}
                \subcaption{$\tilde a_0=\tilde a_1=0.5$, \textbf{exponential} kernel}\label{fig_nonlocal_4c} \includegraphics[width=\textwidth, keepaspectratio=true]{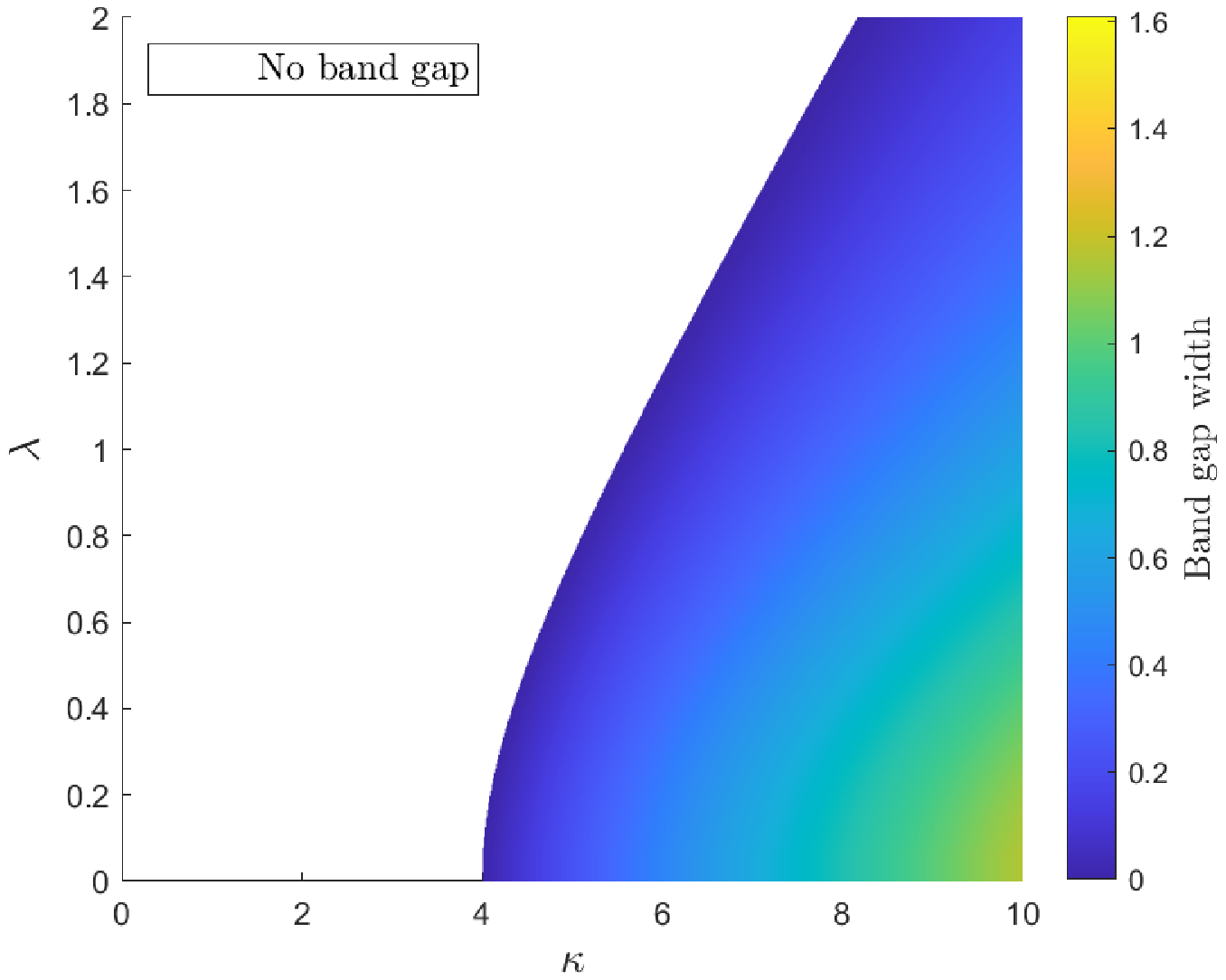}
               
\end{subfigure}
\begin{subfigure}[b]{0.48\textwidth}
               \subcaption{$\tilde a_0=\tilde a_1=0.7$, \textbf{exponential} kernel} \label{fig_nonlocal_4d} \includegraphics[width=\textwidth, keepaspectratio=true]{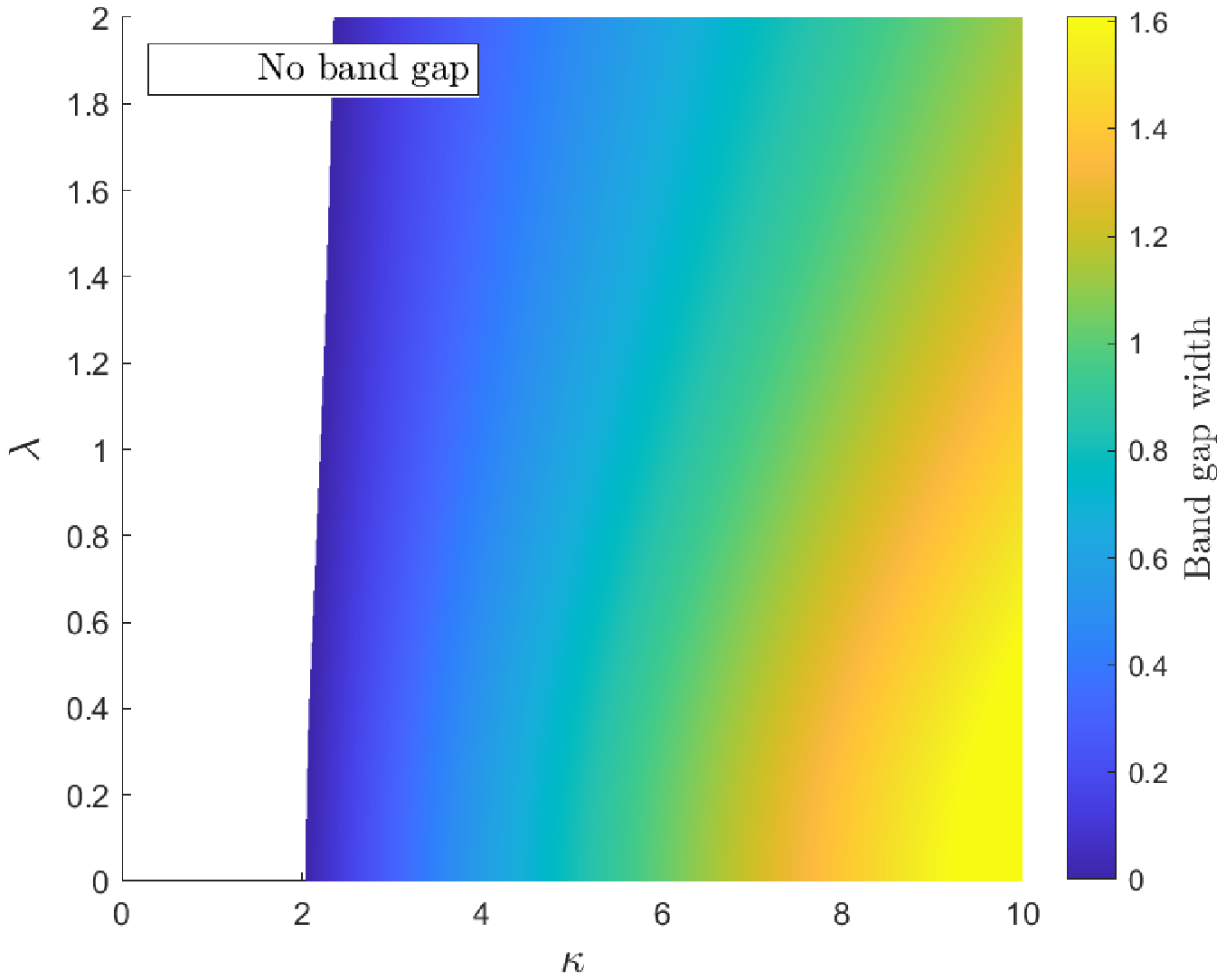} 
                
\end{subfigure}
\begin{subfigure}[b]{0.48\textwidth}
               \subcaption{$\tilde a_0=\tilde a_1=0.5$, \textbf{quartic} kernel}\label{fig_nonlocal_4e} \includegraphics[width=\textwidth, keepaspectratio=true]{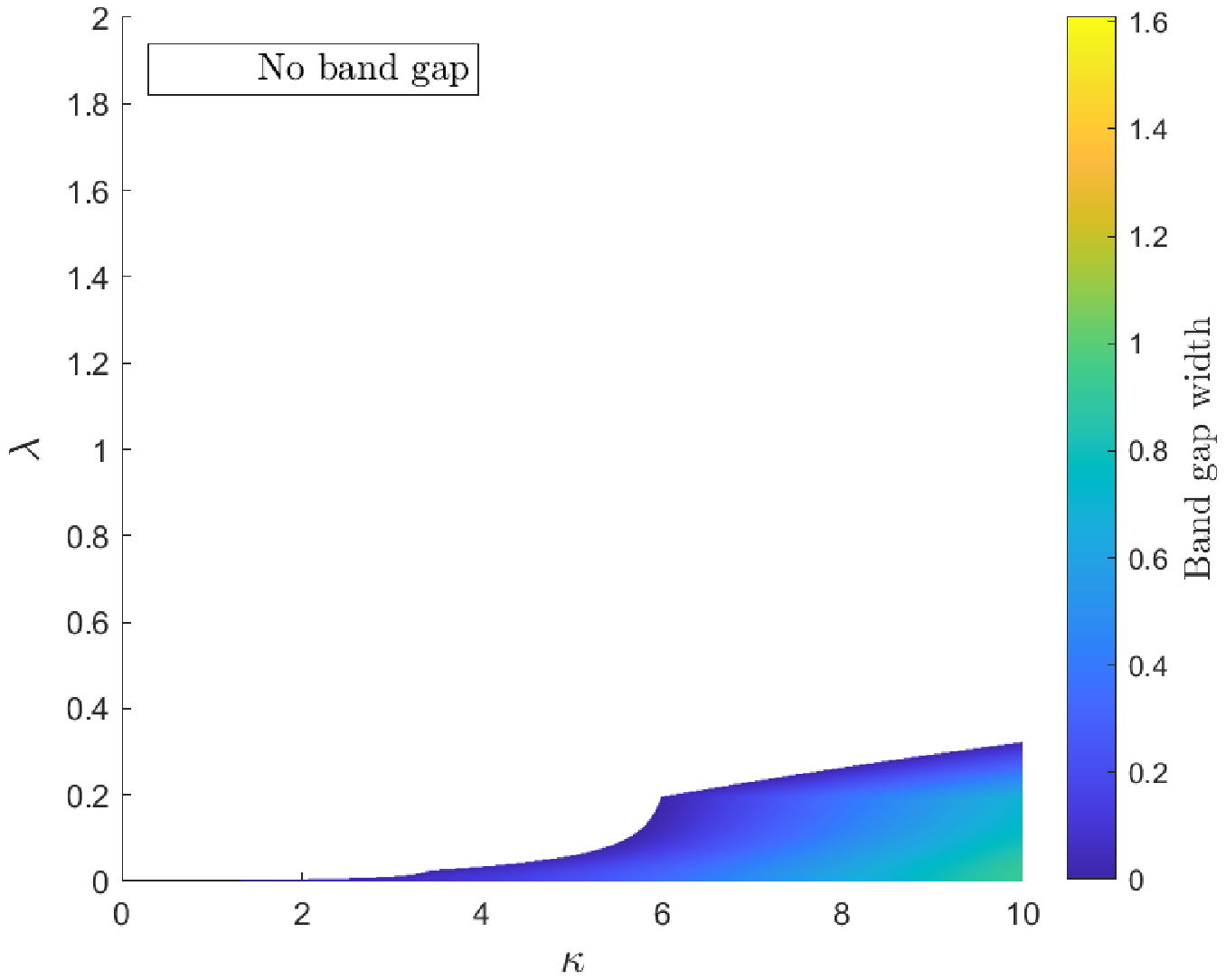}
                
\end{subfigure}
\begin{subfigure}[b]{0.48\textwidth}
               \subcaption{$\tilde a_0=\tilde a_1=0.7$, \textbf{quartic} kernel} \label{fig_nonlocal_4f} \includegraphics[width=\textwidth, keepaspectratio=true]{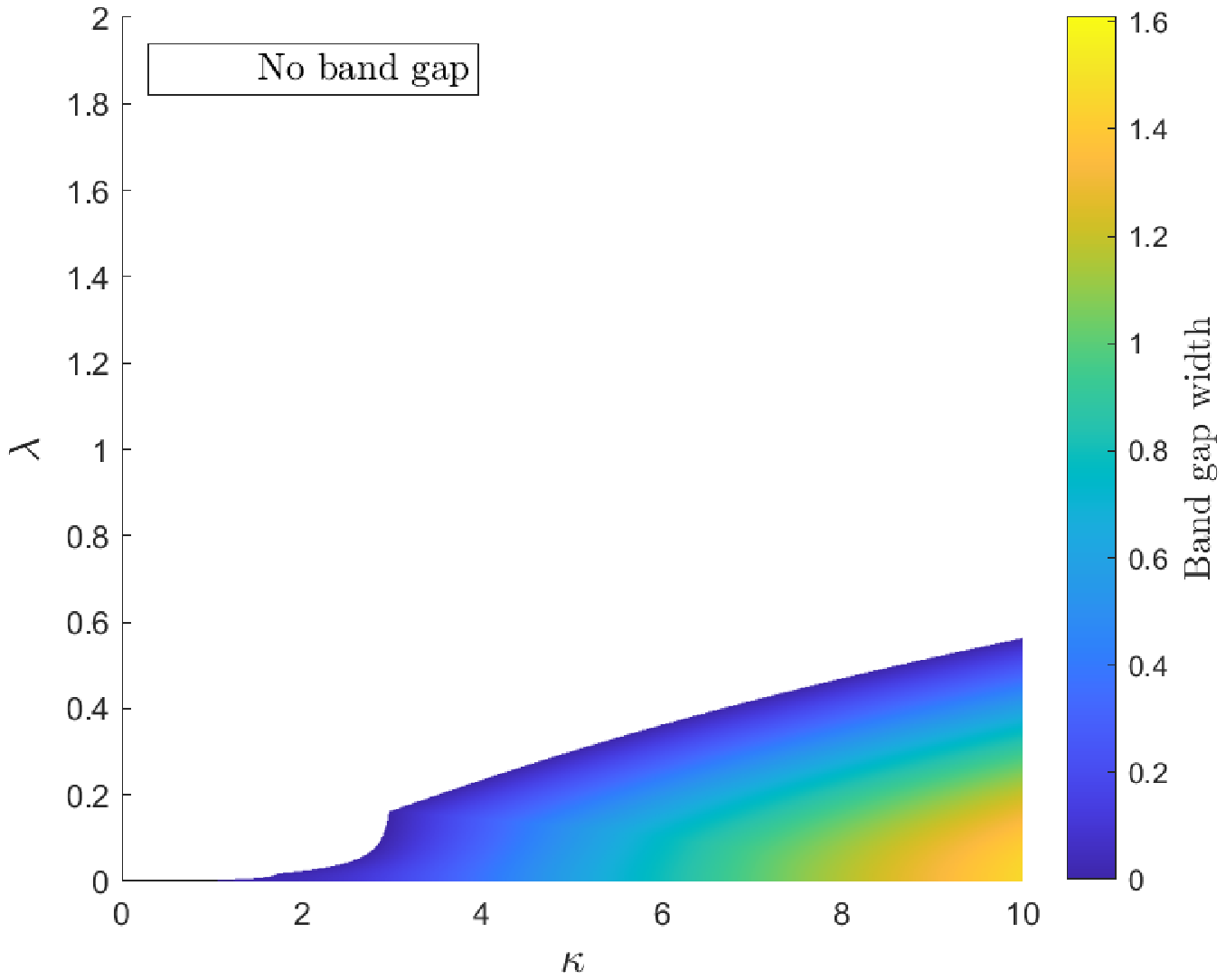} 
                
\end{subfigure}
                \caption{
                The width of the band gap depending on parameters $\kappa$ and $\lambda$ (white region corresponds to combinations of parameters for which no band gap exists).
                }
                \label{fig_nonlocal_4}
\end{figure}

\begin{figure}[h]
    \centering
    \includegraphics[width=0.5\textwidth, keepaspectratio=true]{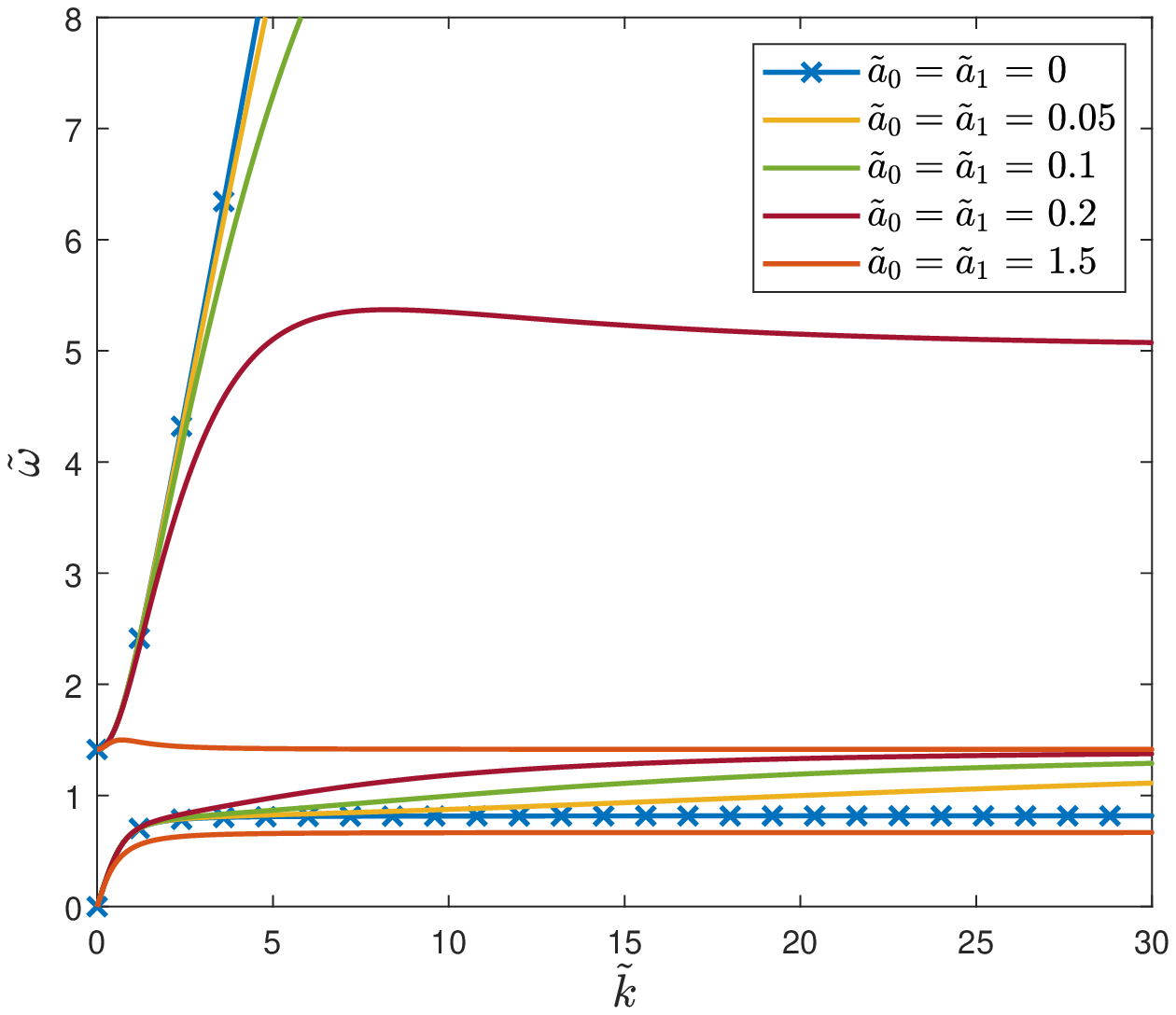}
    \caption{Dispersion curves for exponential kernel with $\lambda=0$ and $\kappa=2$ and various length parameters $\tilde{a}_0=\tilde{a}_1$.}
    \label{fig10}
\end{figure}

\section{Conclusions}
In the paper we demonstrated that, in the one-dimensional setting, a combination of the micromorphic elastic continuum theory and the nonlocal integral averaging approach allows to model two branches of the dispersion curve separated by a band gap in a rather flexible way. In the standard {\bf local} micromorphic case, the band gap appears only when the micromorphic modulus $A$
vanishes, i.e., the kinetic energy is enriched by a term
dependent on the rate of the micromorphic variable, but the potential energy is not enriched by a term dependent on the spatial gradient of the micromorphic variable (it is enriched only by a penalty term that contains parameter $H$ and forces the micromorphic variable to remain close to the local strain). Such a special case might be considered
as somewhat artificial.
In contrast to that, a {\bf nonlocal} extension of the micromorphic elastic continuum permits modeling of
band gaps for a much wider range of parameter combinations, including cases in which the potential energy depends
on the spatial gradient of the micromorphic variable
(i.e., parameter $A>0$). 

The proposed extension introduces nonlocality via two terms present in the expression for potential energy. One is the strain-related term and the other is the penalty term, this time linking the micromorphic variable to the nonlocal strain (instead of the local one). The nonlocal weight functions used in these two terms can be in principle different. Detailed analysis of the dispersion diagram reveals that both extensions are important for the formation of a band gap. The relative strength of nonlocality
is expressed by two dimensionless factors that represent the ratios of the characteristic lengths of the nonlocal averaging operators to the internal length $\ell$ that can be deduced from the ratio between
the micromorphic density and the standard mass density. For a fixed combination of the other model parameters, the nonlocal characteristic lengths
must have a certain minimum value for the band gap to exist,
and the width of the band gap typically increases with
increasing nonlocality. Conversely, if the ratios
between the nonlocal characteristic lengths and the
internal length $\ell$ are fixed, 
a band gap forms only if the micromorphic
stiffness (parameter $A$) does not exceed an upper bound that
depends on the penalty parameter $H$. 
In general, the upper bound for $A$ increases with increasing $H$. 
If the exponential kernel is used, $H$ itself must exceed a certain positive minimum, which is inversely proportional to the square of the nonlocal characteristic length. For the Gaussian and quartic kernels,
it is sufficient if $H$ is positive, but for small values
of $H$ the upper bound for $A$ is also small.
 In the extreme case when
the nonlocal lengths vanish (and thus the model reduces to a local one), $A$ must be strictly zero
for a band gap to exist while $H$ needs to be positive but can be arbitrarily small.

\clearpage
\section*{Acknowledgements}
Financial support received by the first two authors
(MJ and MH)
from the Czech Science Foundation (project No.\ 19-26143X)
is gratefully acknowledged.
The third author (M\v{S}) was supported by the European Regional Development Fund 
(Center of Advanced Applied Sciences, project CZ.02.1.01/\-0.0/\-0.0/\-16\_19/\-0000778)
and by
the Czech Technical University in Prague (internal projects
SGS22/030/OHK1/1T/11 and OHK1-017/23).




 \bibliographystyle{elsarticle-num-names.bst} 
 \bibliography{Reference.bib}





\end{document}